\documentclass[11pt,twoside]{article}
 \usepackage{atmp}
\newtheorem{theorem}{Theorem}[section]
\newtheorem{proposition}[theorem]{Proposition}
\newtheorem{corollary}[theorem]{Corollary}
\newtheorem{lemma}[theorem]{Lemma}
\newtheorem{definition}[theorem]{Definition}

\newcommand{\myendproof}{\hspace*{\fill}{{\bf \small Q.E.D.}}
\vspace{10pt}}

\newcommand{\rd}{{\rm d}}
\newcommand{\bege}{\begin{equation}}
\newcommand{\eeq}{\end{equation}}
\newcommand{\bey}{\begin{eqnarray}}
\newcommand{\eey}{\end{eqnarray}}

\newcommand{\sfrac}[2]{{\textstyle \frac{#1}{#2}}}

\newcommand{\bn}{{\bf n}}

\newcommand{\e}{\varepsilon}

\newcommand{\cU}{{\cal U}}

\newcommand{\bR}{{\bf R}}
\newcommand{\bC}{{\bf C}}

\newcommand{\Tr}{\mbox{Tr}}

\newcommand{\wt}{\widetilde}
\newcommand{\wh}{\widehat}

\newcommand{\cS}{{\cal S}}
\newcommand{\cC}{{\cal C}}
\newcommand{\cF}{{\cal F}}
\newcommand{\cA}{{\cal A}}

\newcommand{\cD}{{\cal D}}

\newcommand{\cK}{{\cal K}}
\newcommand{\cH}{{\cal H}}
\newcommand{\cL}{{\cal L}}

\newcommand{\cO}{{\cal O}}

\newcommand{\no}{\nonumber}

\input epsf

\newcommand{\donothing}[1]{}

\begin{document}

\title{Derivation of the nonlinear Schr\"odinger equation
 from a many body Coulomb system}
\url{math-ph/0111042}
\author{L\'aszl\'o Erd\H os}
\address{School of Mathematics, GeorgiaTech \\ Atlanta, GA 30332}
 \addressemail{lerdos@math.gatech.edu}
\author{Horng-Tzer Yau}
\address{Courant Institute, New York University \\
251 Mercer Str. New York, NY 10012}
\addressemail{yau@cims.nyu.edu}
%\date{Apr 20, 2002}
 \markboth{\it DERIVATION OF NONLINEAR SCHR\"ODINGER EQUATION
\ldots}{\it L. ERD\H OS,
H.-T. YAU}

%\date{Apr 20, 2002}

%\maketitle

\begin{abstract}
We consider the time evolution of $N$ bosonic particles
interacting via a mean field Coulomb potential. Suppose the initial state
is a product wavefunction. We show that at any finite time the correlation
functions factorize in the limit $N \to \infty$. Furthermore, the limiting
one particle density matrix satisfies the nonlinear Hartree equation. The
key ingredients are  the uniqueness of the BBGKY hierarchy
for  the correlation functions  and a new apriori estimate for the
many-body Schr\"odinger equations.
\end{abstract}



%\tableofcontents

\section{Introduction}

It is well-known that the Schr\"odinger equation  governs the
dynamics of non-relativistic quantum systems. The basic existence
and uniqueness of the Schr\"odinger equation were studied
extensively for many systems, including the important  Coulomb
systems. Despite these efforts, there are few cases that the
solutions to the many-body Schr\"odinger equation can be
described in reasonable  details. One such case is a system of $N$
weakly interacting bosons, or more precisely, $N$ bosons
interacting with a pair potential whose coupling constant is of
order $N^{-1}$.

For such a system, we expect that the potential acting on any
single particle is approximately  generated by the
average density of
all other particles. Therefore, there are no correlations
generated by the dynamics and all correlations of the state at
a later time are almost all due to the initial correlations. In the
simplest scenario, i.e.,  when
the initial state has no correlation, the
correlations at a later time should be negligible as well. We consider
initial states  of $N$ bosons
that are described by products of one-particle wave functions.
Therefore, our goal is to
show that, if the initial $N$-body  wave function is a product of
a one-particle wave function, then
it is well-approximated by a  product of some one-particle wave
function at later times as well.
Assuming this is correct, it is easy to see that the one
particle wave function satisfies a nonlinear Schr\"odinger
equation whose nonlinearity is given by the convolution of the
potential with the mass density of the one particle wave function,
i.e., $V \ast |\psi|^2$. This equation is typically called the
{\it Hartree equation}. In the special case when the pair interaction is
Coulomb, it is called the Schr\"odinger-Poisson system. As usual
for the Schr\"odinger equation, we can always cast it in the form
of an equation for the density matrix.
We shall use the same name for both setups.

This problem has a long history and has been considered by many
authors. For regular pair-potentials, it was solved by Hepp
\cite{H} in the context of field operators and by Spohn \cite{Sp}
using density matrices. Hepp's method requires differentiability
of the potential, while Spohn uses only the boundedness of the
potential.  Ginibre and Velo \cite{GV-1} have greatly
extended these results to  include singular potentials (including
Coulomb). However they worked in the second quantized framework
and the quasi-free initial state used in \cite{GV-1} describes
special classes of excitations around the product state. In particular, this
method does not apply to wave-functions with a fixed number of particles.

The basic tool to analyze the many-body dynamics is the BBGKY
hierarchy. We notice that the time derivative of the $k$-particle
correlation function (more precisely, $k$-particle density
matrix) of an $N$-body state is given by the $k+1$ particle
correlation function for each
$k=1, 2, \ldots N-1$. This gives a
hierarchy of coupled evolution equation for all $k$ correlation
functions up to $N$, called the {\it Schr\"odinger hierarchy}. Without
further limiting procedure, this system is just a rewriting of the
original Schr\"odinger equation. If we take  the {\it formal}
$N\to \infty$ limit,  it converges to an infinite system of equation,
called the {\it BBGKY hierarchy}, for the $k$-particle density
matrices for $k=1, 2, \ldots $. For a product initial state this
infinite hierarchy has a trivial solution of product form built up
from the solution of the Hartree equation. Although the $N$-body
hierarchy has a unique solution, the infinite hierarchy may, in
principle, admit more than one solution. Thus our task is to
establish the convergence of the Schr\"odinger hierarchy to the
BBGKY hierarchy as $N\to \infty$, and prove the uniqueness
 of the solution to the BBGKY hierarchy.

Both the BBGKY and the Schr\"odinger hierarchy can be put into
systems  of integral equations.  For bounded potential,  these
systems can be solved by iteration which converges in the trace
norm of the density matrices. Using this idea on the Schr\"odinger
hierarchy, Spohn proved that the $k$-point correlation functions
of the $N$ body system are given by  convergent power series with
an error estimate uniformly in $N$. Thus one can take  the limit
$N \to \infty$ and one obtains that the $k$-point correlation
functions are indeed given by the products of the solution to the
nonlinear Hartree equation.

Bardos et. al. have followed a  different route \cite{BGM}. They
showed that any $w^*$-limit point of the solutions to the $N$-body
hierarchy satisfies the infinite BBGKY hierarchy. This proof
requires the potential to be bounded below and  in $L^2 +
L^\infty$. If, in addition, the potential is bounded, then the
sequence of correlation functions is shown to be $w^*$-precompact in
the trace norm. The uniqueness of the  solution to the infinite
hierarchy is established by controlling the trace norm similarly
to \cite{Sp}. In particular, this result establishes the
convergence to the BBGKY hierarchy for the  repulsive Coulomb
systems which was not covered in  Spohn's work.

%Compactness and uniqueness implies that the sequence of $N$-body
%solutions converges to the unique solution of the BBGKY hierarchy.

The main question is the uniqueness of the BBGKY hierarchy in the
case of singular potentials. Furthermore, the convergence in the
attractive Coulomb case should be resolved since it describes the
important  gravitational system. Technically, all the uniqueness
methods  rely heavily on the boundedness of the potential via the
estimate $\Tr | V\gamma| \leq \| V\|_{\infty} \Tr|\gamma|$ for a
density matrix $\gamma$.  The Coulomb potential, i.e., $V(x) =\pm
|x|^{-1} $,  is unbounded, but Hardy's inequality allows us to
control it in $H^1$-norm, i.e.,
$$
\Tr \; \big|  V \gamma \big| \leq C\big(\; \Tr \; \nabla \gamma \nabla +
\Tr \; \gamma \;\big)\; .
$$

Therefore the solution to the
BBGKY hierarchy should be unique under the right Sobolev
norm. Since we have correlation functions of arbitrary number of
particles, the Sobolev space we choose is the iterative Sobolev
space (see Section \ref{sec:sob} for the precise definition)
which is weaker than the usual one but sufficient for the
uniqueness. This poses the problem of estimating the iterative
Sobolev norm for correlation functions. This estimate can only be
obtained from the original $N$-body Schr\"odinger equation for which
we know only the conservation of mass and energy. Notice that in
\cite{Sp} and \cite{BGM} the only  estimates obtained are
trace norm and  $H^1$ norm bounds, which are  consequences of
the conservation of mass and energy for the  $N$-body Schr\"odinger
equation.

Since we can only establish the uniqueness of the BBGKY hierarchy
in the iterative Sobolev norm, we need to establish an apriori
estimate that the $k$-correlation functions are bounded in such a
norm. It should be emphasized that to establish such estimates
directly for the correlation functions of a Coulomb system is very
difficult due to the $|x|^{-1}$ singularity. However, it is feasible
for a cutoff Coulomb systems with an $N$ dependent cutoff. The
main idea here is to take the energy to a higher power and
deduce from there the estimate on the iterative Sobolev norm.
Notice that the third power of the potential is already not in
$L^1$.  Hence it does not even define a meaningful operator in the
usual sense. This is part of the technical reason we need to
perform a truncation for the Coulomb singularity.

We now need to control the original evolution and the cutoff
dynamics. Here we use the conservation of the $L^2$ norm to
control the difference of these two evolutions.  Since to control
an $N$-body system in general produces a factor $N$, our cutoff
has to be rather small. The various restrictions on the cutoff
scale finally give the choice  of order $o(N^{-1/2})$.

\medskip

This work was partly inspired by the work of \cite{BGM}
and partly by the question posed by J. Yngvason regarding the derivation
of  the time dependent
Gross-Pitaevskii equation
from the many-body Schr\"odinger equation. Unfortunately,
our method, as it stands,  still cannot prove the convergence to the
Gross-Pitaevskii equation.
Another possible motivation for studying high density bosons
with Coulomb interaction is that electrically
charged ions may have bosonic statistics.

This work was supported by NSF grants
DMS-9970323 (L.E.) and DMS-0072098 (H.-T. Y.)
The discussions with C. Bardos,   F. Golse, N. Mauser
and J. Yngvason are gratefully acknowledged. The results of this work
and \cite{BGM} were jointly outlined in \cite{BEGMY}.

\section{Many-body Schr\"odinger evolution with mean field potential}
\setcounter{equation}{0}

Consider a system of $N$-bosons weakly interacting with a Coulomb
potential. The dynamics of such a system is governed by the
Schr\"odinger equation with the Hamiltonian
$$
    H_N = -\frac{1}{2} \sum_{\ell=1}^N \Delta_{x_\ell}
    + \frac{1}{N} \sum_{\ell<j} V( x_\ell-x_j)
$$
defined on $\bigotimes_{\ell=1}^N L^2(\bR^3)$, where $V(x)= \pm \mu
|x|^{-1}$ with $\mu>0$.
The wave functions are symmetric functions
of $N$ variables. Since the Hamiltonian is symmetric, the wave function
at the time $t$ will be symmetric as long as the initial data is symmetric.

The Schr\"odinger equation is given by
\begin{equation}\label{N-Sch}
i \partial_t \Psi_{N,t} = H_N \Psi_{N,t},
\end{equation}
with the initial data specified at the time $t=0$.
The equation can be solved explicitly by  $\Psi_{N,t} =
e^{-itH_N}\Psi_{N,0}$. Let $\gamma_{N,t} = \pi_{\Psi_{N,t}}$ be the
projection operator in $L^2(\bR^3)$ associated with the wave function
$\Psi_{N,t}$. The Schr\"odinger equation  is equivalent to
the operator equation
\begin{equation}\label{N-Sch-op}
i \partial_t \gamma_{N,t} = [ H_N,  \gamma_{N,t}]\;.
\end{equation}
This is a more general setup since it allows  a general density
matrix not coming from a wave function.
The $n$-point density matrix of $\gamma_{N, t}$ is defined by
\bey
 \gamma^{(n)}_{N, t} (x_1, \ldots x_n; x_1', \ldots x_n'): =
\int \Psi_{N,t} (x_1, \ldots x_n, x_{n+1}, \ldots x_N)\qquad\qquad\qquad\\
    \times\overline{\Psi_{N,t}} (x_1', \ldots x_n', x_{n+1}, \ldots x_N)
    \rd x_{n+1}\ldots \rd x_N \no
\eey
if $n\leq N$ and  $\gamma^{(n)}_{N, t} :=0$ otherwise.
The normalization is $(n\leq N$)
\begin{equation}
    \Tr \; \gamma^{(n)}_{N, t} = 1 \;.
\label{trace1}
\end{equation}

It is a simple calculation that the
$n$-point density matrices of the solution to
the  Schr\"odinger equation satisfy the finite hierarchy
\bey \label{eq:finBBGKY}
\lefteqn{\gamma_{N,t}^{(k)}(x_1, \ldots x_k;
    x_1', \ldots x_k')}  \\
    & = &  \cU_{N,k}(t)\gamma_{N,0}^{(k)}
    (x_1, \ldots x_k;
    x_1', \ldots x_k')
 + (-i)\sum_{\ell=1}^k \int_0^t \rd s \; \cU_{N,k}(t-s) \no \\
& &     \frac{N-k}{N}\Bigg[ \int \rd x_{k+1}
    \Big( V(x_\ell-x_{k+1})- V(x_\ell'-x_{k+1}) \Big)\no\\
   && \times  \gamma_{N,s}^{(k+1)}(x_1, \ldots x_k, x_{k+1};
    x_1', \ldots x_k', x_{k+1})\Bigg] \no
\eey
where $\cU_{N,k}(t)\gamma = e^{-itH_{N}^{(k)}}\gamma e^{itH_{N}^{(k)}}$
and
$$
    H_{N}^{(k)} =- \frac{1}{2}\sum_{\ell=1}^k \Delta_\ell
    + \frac{1}{N} \sum_{\ell<j}^k V(x_\ell-x_j) \; .
$$
If we take the limit $N \to \infty$ and neglect all lower
order terms, this system converges to the BBGKY hierarchy
given by the following infinite system of equations
for density matrices
\bey\label{eq:BBGKY}
   \lefteqn{ \gamma_{t}^{(k)}(x_1, \ldots x_k;
    x_1', \ldots x_k' )}\qquad\qquad\\
    & =&  \cU_k(t)
    \gamma_{0}^{(k)}(x_1, \ldots x_k;
    x_1', \ldots x_k' )
     + (-i)\sum_{\ell=1}^k \int_0^t \rd s \; \cU_k(t-s) \no \\
 &&   \Bigg[ \int\Big( V(x_\ell-x_{k+1})- V(x_\ell'-x_{k+1}) \Big)\no\\
    &&\times \gamma_{s}^{(k+1)}
    (x_1, \ldots x_k, x_{k+1}; x_1', \ldots x_k';, x_{k+1})\rd x_{k+1}
    \Bigg] \no
\eey
where $\cU_k(t)\gamma :=  \Big(\prod_{j=1}^k e^{-it\Delta_j/2}\Big)
\; \gamma \; \Big(\prod_{j=1}^k e^{-it\Delta_j/2}\Big)$, $k=1,2,\ldots$.
This is the integral form of an infinite system of evolution
equations with initial data $\Gamma_0 :
 = \big( \gamma^{(1)}_0, \gamma^{(2)}_0,
\ldots \big)$.

\bigskip

Suppose the initial data is of product form and given by
\begin{equation}\label{initial}
\Psi_{N,0} (x_1, \ldots x_N) : = \prod_{j=1}^N \psi_0(x_j)
\end{equation}
for some initial one-particle wave function $\psi_0$. We always assume
that $\psi_0$ is normalized in $L^2(\bR^3)$, i.e.  $\|\psi_0\|_2 =1$.
For this initial data,
the $k$-point density matrix, $\gamma_{N, 0}^{(k)}$ is, simply the
tensor product $\bigotimes_{1}^k
\gamma_{\psi_0}$
 of the one-point density matrix $ \gamma_{\psi_0}$,
which is the projection matrix onto $\psi_0$.
 One can check that
if the initial data to the BBGKY hierarchy is of product form, then
there is a special solution to the BBGKY hierarchy which is of product
form and the one-particle density matrix  satisfies
the nonlinear Schr\"odinger equation
\begin{equation}
    i\partial_t \gamma_t =  \Big [ \; -\frac{1}{2}\Delta_x
    +\Big(\int V(\cdot -z)\gamma_t(z, z) \rd z\Big)\, , \; \gamma_t
    \; \Big ] \; .
\label{eq:NLS-op}
\end{equation}
If we denote the kernel by
$\gamma_{t}(x, x')$, then
(\ref{eq:NLS-op}) is equivalent to
\bey\label{eq:NLSgam}
\lefteqn{i\partial_t\gamma_{t}(x, x')
    = -\frac{1}{2} \Big[\Delta_x - \Delta_{x'}\Big]\gamma_{t}(x, x')}
    \qquad\qquad\;\\
    &&+ \int \rd z\Big[ V(|x-z|)-V(x'-z)\Big]\gamma_{t}(z,z)
    \gamma_{t}(x, x') \; .\no
\eey
We can put this equation into a more familiar form. If
$\gamma_{0}(x, x') : = \psi_0(x)\overline{\psi_0(x')}$,
then $\gamma_{t}(x, x') : = \psi_t(x)\overline{\psi_t(x')}$
where $\psi_t$ satisfies
\begin{equation}
    i\partial_t \psi_t(x) = -\frac{1}{2}\Delta_x\psi_t(x)
    + \Big(\int V(x-z)|\psi_t(z)|^2 \rd z\Big)\psi_t(x)
\label{eq:NLS}
\end{equation}
with initial data $\psi_{t=0}=\psi_0$.
This equation was studied extensively by Ginibre and Velo.
In particular, the equation preserves the $L^2$ norm and
the energy and we have \cite{GV-2}
\begin{equation}
    \sup_{ t \ge 0} \| \psi_t \|_{H^1} <\infty
\label{eq:GV}
\end{equation}
if the $H^1$ norm of the
initial condition $ \|\psi_0 \|_{H^1}$ is finite.
Although the equation is nonlinear, the normalization
condition $\|\psi_0\|_2=1$ can be assumed without loss
of generality at the expense of changing $\mu$.

Therefore, if we can justify the limiting procedure and
prove the uniqueness of the BBGKY hierarchy, the evolution
of the weakly interacting $N$-bosons can be understood
by a one-body nonlinear Schr\"odinger equation. We first
describe the topology for the limiting procedure.

Denote by $H$  the Hilbert space $L^2(\bR^3)$ and let $\cL(H)$ be
the set of bounded operators with operator norm $\|\cdot\|$. We let
$\cL^1(H)$ be the set of trace class operators with the
norm $\|\gamma\|_1 : = \Tr |\gamma|$. The set of density matrices,
$\wh\cL^1$, is defined as the subset of nonnegative self-adjoint trace
class operators.
Let $\cK:=\cK(H)$
be the set of compact operators equipped with the operator norm.
It is well-known  (Theorem VI.26 in Vol.I. of \cite{RS})
 that the dual space of the compact operators is
the space of trace class operators, i.e.,
$(\cK, \| \cdot \|)^* =
(\cL^1, \| \cdot \|_1)$. This gives rise to
the variational characterization of the trace norm:
\begin{equation}
    \| A \|_1 = \sup_{K\in \cK(H)\; : \;\| K\|=1} \Big| \; \Tr \; AK\;
     \Big| \; .
\label{eq:var}
\end{equation}
The $w^*$-topology on $\cL^1$ is induced by a family of seminorms
$A\to |\Tr \; AK |$ that are indexed by the family of compact
operators $K\in \cK$.

Denote by $H^{\otimes k} : = \bigotimes_{i=1}^k H =
\bigotimes_{i=1}^k L^2(\bR^3)$ the $k$-tensor product of $L^2(\bR^3)$.
We can define the trace norm on $H^{\otimes k}$
and extend the duality from $k=1$ to all $k$.
Define the space
$$
    \cC: =
    \Big\{ \Gamma = (\gamma^{(1)}, \gamma^{(2)}, \ldots ) \; : \;
    \gamma^{(k)} \in  \cL^1(H^{\otimes k}) \Big\}
    = \prod_{k=1}^\infty \cL^{1}(H^{\otimes k}) \; .
$$
We equip this set  with the product of the $w^*$-topologies
on each component. Thus the convergence in $\cC$ is characterized by
the following property: $\Gamma_n\to \Gamma$
as $n\to\infty$,
if  for each $k$ we have $\gamma^{(k)}_n \to  \gamma^{(k)}$
in $w^*$ sense in  $\cL^{1} (H^{\otimes k})$.

Let $\langle t \rangle : = (1+t^2)^{1/2}.$
We define the set
$$
    L^\infty(\bR_+, \langle t \rangle^{-1} \rd t,  \cC): =
    \prod_{k=1}^\infty L^\infty\Big(\bR_+,  \langle t \rangle^{-1} \rd t,
    \cL^{1}(H^{\otimes k})\Big)
$$
i.e., the set of functions $\gamma^{(k)}(t):
\bR_+ \to \cL^1(H^{\otimes (k)})$ with $\sup_t \langle t \rangle^{-1}
\|\gamma^{(k)}(t)\|_1 <\infty$.
We equip this set with the product of the $w^*$-topologies
on each factor. On each factor  the $w^*$-topology is
given by seminorms
$$
    \gamma^{(k)} (t) \to \Big| \int_0^\infty \Tr \Big[ \;
    K(t)\gamma^{(k)}(t) \;\Big] \; \rd t
     \Big|
$$
where $K(t)\in L^1(\bR_+, \langle t \rangle\rd t,
 \cK(H^{\otimes k}))$, i.e. $\int_0^\infty
 \| K(t)\| \langle t \rangle \rd t <\infty$.

Let
\begin{equation}
    \Gamma_{N,t} : = \Big( \gamma_{N,t}^{(1)}, \gamma_{N,t}^{(2)},
    \ldots \Big) \in \cC\; .
\label{def:Gamma}
\end{equation}
For any one-particle wave function $\psi$ we define
$$
    \gamma^{(n)}_{\psi} (x_1, \ldots x_n; x_1', \ldots x_n')
    : = \prod_{j=1}^n \gamma_{\psi} (x_j, x_j')\; .
$$
Suppose $\psi_t \in H^1(\bR^3)$ is a solution to the nonlinear
Schr\"odinger equation  (\ref{eq:NLS}), then the collection of
density matrices
\begin{equation}
    \Gamma_{\psi_t} : = \Big( \gamma_{\psi_t}^{(1)},
    \gamma_{\psi_t}^{(2)},
    \ldots \Big) \in \cC
\label{Gammapsi}
\end{equation}
is a solution to (\ref{eq:BBGKY}).
We now state the main theorem:

\begin{theorem}\label{thm:main}
Let $V(x)= \pm\mu |x|^{-1}$ be the repulsive or attractive Coulomb
potential with some $\mu>0$.
Assume that $\psi_0 \in H^2(\bR^3)$ and let $\psi_t$ be the solution
of (\ref{eq:NLS}). Let $\Gamma_{N,t}$ be the solution to
(\ref{eq:finBBGKY}) with initial condition $\Gamma_{N,0}: = \Gamma_{\psi_0}$.
As $N\to\infty$, we have
$$
    \Gamma_{N,t} \to \Gamma_{\psi_t}
$$
in the w* topology of  $L^\infty(\bR_+,\langle t \rangle^{-1} \rd t, \cC)$.
 In other words, for each $k$
$$
    \gamma^{(k)}_{N, t} \to \gamma^{(k)}_{\psi_t}
$$
in the weak* topology of $L^\infty(\bR_+,\langle t \rangle^{-1} \rd t,
 \cL^{1}(H^{\otimes k}) )$.
\end{theorem}

\noindent
{\it Remark 1:} Note that our theorem requires the initial  one-particle
wave function
to be in $H^2$ although the natural space for the nonlinear
Schr\"odinger equation (\ref{eq:NLS}) is $H^1$.

\medskip

\noindent
{\it Remark 2:} Our proof works for more general potentials as well,
for example one can consider $V(x) = A(x) |x|^{-1} + B(x)$
with $A, B\in \cS(\bR^3)$ Schwarz class. The control on high
derivatives is necessary because of the commutator estimate (\ref{eq:Uijest}).

\bigskip

We close this section by an important remark.
The equations of the BBGKY hierarchy
are  written for the kernels of the density matrices,
however  we interpret them  as equations for density matrices
even though we sometimes use the traditional kernel notation.
This point of view allows us to circumvent the
problem that apriori the kernels are only functions
defined almost everywhere, hence setting $x_{k+1}' = x_{k+1}$ in
the interaction term requires an extra argument.
In this paper we interpret the integral
$$
    \int V(x_\ell - x_{k+1}) \gamma^{(k+1)} (x_1, \ldots x_{k}, x_{k+1};
     x_1', \ldots x_k';, x_{k+1})\rd x_{k+1}
$$
as the partial trace
$$
    \Tr_{x_{k+1}} \wt V\gamma^{(k+1)}
$$
where $\wt V$ is the multiplication operator by $ V(x_\ell - x_{k+1})$.
We will always verify that $\wt V \gamma^{(k+1)}$ is a trace
class operator on $L^2(\bR^{3(k+1)})$, hence the
partial trace is well defined. In the Appendix we
give the precise definition of the partial trace and
we collect a few basic information.

\bigskip

\noindent
{\it Convention:} Throughout the paper the letter $C$ will refer
to various constants depending only on $\mu$.

\section{Sobolev spaces of density matrices}\label{sec:sob}
\setcounter{equation}{0}

We recall that
$\cL(H)$, $\cL^1(H)$ and $\cK(H)$ denote  the set of
bounded, trace class and compact operators on the Hilbert space $H$, respectively.
Also,  $\wh\cL^1\subset \cL^1(H)$ denote the set of density matrices.
We have the standard inequality
\begin{equation}
    \| AB \|_1 \leq \| A\| \; \| B\|_1
\label{eq:AB}
\end{equation}
and  the variational characterization of the trace norm
(\ref{eq:var}).
We also recall that $|A+B|\leq |A| + |B|$ is {\it not} true in general,
but $\Tr |A+B| \leq \Tr |A| + \Tr |B|$ is true, following
from (\ref{eq:var}), (\ref{eq:AB}). Moreover,
if we have two Hilbert spaces, $H_1$, $H_2$, then
the following inequality is valid for partial traces
\begin{equation}
    \Tr_1 \; \Big| \; \Tr_2 \; A \; \Big| \leq \Tr_{1,2}\;  |A| \; .
\label{eq:fubini}
\end{equation}
In the Appendix we give the proof of this
inequality and we also collect a few basic facts about partial
traces.

We define the analogue of the
$H^1$ Sobolev norm for trace class
operators as follows:
$$
    \| T \|_{\cH^1}: = \Tr \;|ST S|
$$
where $S:= (I-\Delta)^{1/2}$.
We introduce the space of  operators $\cH^1= \cH^1(H)$ as:
$$
    T \in \cH^1 \Longleftrightarrow \| T \|_{\cH^1}< \infty
$$
and let $\wh\cH^1: = \cH^1 \cap \wh\cL^1$. Notice that
$\|\gamma\|_{\cH^1} = \Tr \; S\gamma S = \Tr\gamma +
\Tr (-\Delta)\gamma$ for $\gamma \in \wh\cH^1$.

We  define a weak* topology on $\cH^1$ and for this
purpose we identify this space with a dual space.
We let
$$
    \cA : = \Big\{ SKS \; : \; K\in \cK\Big\}
$$
and we equip this space with the norm
$$
    \| T \|_{\cA} : = \| S^{-1} T S^{-1}\| \; .
$$

\begin{lemma}\label{lemma:dual}
With the notations above
$$
    ( \cA, \| \cdot \|_\cA)^* = (\cH^1, \| \cdot \|_{\cH^1}) \; .
$$
\end{lemma}
The proof of this Lemma is found in the Appendix
and it is similar to the standard proof of
$(\cK, \|\cdot\|)^*= (\cL^1, \|\cdot \|_1)$
outlined in \cite{RS}.
Using this duality, we equip $\cH^1$ with a $w^*$-topology,
induced by the seminorms  $T \to | \Tr \; AT |$
 indexed by elements $A\in\cA$.

\begin{lemma}\label{lemma:wc}
Let $\gamma_n \in \wh \cH^1$ be a sequence of uniformly
bounded density matrices, i.e.
$$
    C:=\limsup_n \|\gamma_n \|_{\cH^1}
    = \limsup_n \Tr \; S\gamma_n S < \infty \; .
$$
Then one can extract a $w^*$-convergent
subsequence, $\gamma_{n_k}\to \gamma$,
$\gamma \in \wh\cH^1$, and any $w^*$-limit point $\gamma$ of
the sequence $\{\gamma_n\}$
 satisfies $\| \gamma \|_{\cH^1}=
\Tr \; S\gamma S \leq C$
\end{lemma}

The convergence in Lemma \ref{lemma:wc} follows from standard
application of the Banach-Alaouglu theorem.
The positivity of the limit
can be checked by testing with
the projection operators $P_f: =|f\rangle\langle f|\in\cA$
for all $f\in H$. \myendproof

\bigskip

\noindent
Define the following norms for operators
$\gamma^{(k)} \in \cL (H^{\otimes k})$
$$
    \| \gamma^{(k)} \|_{\cH^{1,(k)}}:
    = \Tr \Big| S_{x_1} \ldots S_{x_k} \gamma^{(k)}  S_{x_k}\ldots S_{x_1}
    \Big| \; .
$$
This norm is equivalent to the norm
$$
    \Tr |\gamma^{(k)}|
    + \Tr \Big| \nabla_{x_1}\nabla_{x_2}\ldots  \nabla_{x_k}
     \gamma^{(k)} \nabla_{x_k} \ldots  \nabla_{x_1}\Big|
$$
and for density matrices we have the identity
$$
    \| \gamma^{(k)} \|_{\cH^{1,(k)}} =
    \Tr (I-\Delta_{x_1}) (I-\Delta_{x_2})\ldots  (I-\Delta_{x_k})
    \gamma^{(k)}\; .
$$
Here we denote by $\cH^{1,(k)} = \cH^{1}(H^{\otimes k})$
the set of operators with finite $\cH^{1,(k)}$ norm.
Since higher derivatives on the same variable are not allowed,
these norms are weaker than the operator analogue
of the traditional higher order $W^{k,1}$ Sobolev norms.

The $w^*$-topology  on the space $\cH^{1,(k)}$ is given by the
seminorms indexed by the set
$$
    \cA^{(k)} : = \Big\{  S_{x_1} \ldots S_{x_k} \;
    K \; S_{x_k} \ldots S_{x_1} \; : \; K\in \cK(H^{\otimes k}) \Big\}
$$
with norm
$$
    \| T\|_{  \cA^{(k)}}:= \Big\| ( S_{x_1} \ldots S_{x_k})^{-1}
    T ( S_{x_1} \ldots S_{x_k})^{-1} \Big\|
$$
and
$$
    \big(  \cA^{(k)}, \| \cdot \|_{  \cA^{(k)}}\big)^* =
    \big( \cH^{1, (k)}, \| \cdot \|_{\cH^{1, (k)}}\big)
$$
analogously to the one variable case in Lemma \ref{lemma:dual}.
The analog of Lemma \ref{lemma:wc} also holds.

\bigskip

Define the set
$$
    \cD: =
    \Big\{ \Gamma = (\gamma^{(1)}, \gamma^{(2)}, \ldots ) \; : \;
    \gamma^{(k)} \in  \cH^1(H^{\otimes k}) \Big\}
    = \prod_{k=1}^\infty \cH^{1}(H^{\otimes k}) \; .
$$
This set is equipped with the product topology $\tau$ generated by
the $w^*$-topology on each $\cH^{1}(H^{\otimes k})$.
The convergence in $\cD$ is characterized by
the following property: $\Gamma_n\to \Gamma$ in $(\cD,\tau)$
as $n\to\infty$,
if  for each $k$ we have $\gamma^{(k)}_n \to  \gamma^{(k)}$
in $w^*$ sense in  $\cH^{1,(k)}$.

\bigskip

Finally, we will consider time dependent density matrices.
The set $ L^\infty(\bR_+,  \langle t \rangle^{-1} \rd t,
 \cH^{1, (k)})$ consists of functions $\gamma^{(k)}(t): \bR_+\to
\cH^{1, (k)}$ with
$\sup_t \langle t \rangle^{-1} \|\gamma^{(k)}(t) \|_{\cH^{1, (k)}} < \infty$.
We define the set $L^\infty(\bR_+, \langle t \rangle^{-1} \rd t,  \cD) =
\prod_{k=1}^\infty L^\infty(\bR_+,  \langle t \rangle^{-1} \rd t,
 \cH^{1, (k)})$ and we equip it with the product of the $w^*$-topologies
on each factor. On each factor  the $w^*$-topology is
 given by seminorms
$$
    \gamma^{(k)} (t) \to \Big| \int_0^\infty \Tr \Big[ \;
    A(t)\gamma^{(k)}(t) \;\Big] \; \rd t
     \Big|
$$
where $A(t)\in L^1(\bR_+, \langle t \rangle\rd t,
 \cA^{(k)})$, i.e. $\int_0^\infty
 \| A(t)\|_{ \cA^{(k)}} \langle t \rangle \rd t <\infty$.

\bigskip

\section{Structure of the proof}
\setcounter{equation}{0}

In this section we give the main propositions and show how
our theorem follows from them. The proofs of the propositions
are found in the subsequent sections.

\subsection{Cutoff of the Coulomb singularity.}\label{sec:cut}

\bigskip

We now introduce an $N$ dependent cutoff for the Coulomb
potential. Decompose  $V=V_1 + V_2$ where
$$
    V_1(x) = \theta (\sqrt{N} \e^{-1}x) V(x)
$$
with a smooth cutoff function $0\leq \theta\leq 1$ with $\theta \equiv 1$ on
outside of $B(0,2)$ and $\theta \equiv 0$ inside $B(0,1)$. Here
$B(x,r)$ denotes the ball of radius $r$ centered at $x$. Notice the
potentials, $V_1$ and $V_2$, depend
 on $\e$ and $N$ which dependence is not explicitly
labelled.

We define the cutoff Hamiltonian
$$
    H_{N,\e} = -\frac{1}{2} \sum_{\ell=1}^N \Delta_{x_\ell}
    + \frac{1}{N} \sum_{\ell<j} V_1( x_\ell-x_j)
$$
and the remaining part of the potential
$$
    W:= W_{N, \e}= \frac{1}{N} \sum_{\ell<j} V_2( x_\ell-x_j)
$$
so that $H_N=  H_{N,\e} + W$.

\bigskip

The density matrices $\gamma_{N, t}^{\e, (k)}$ of the
cutoff dynamics again satisfy the $N$-body Schr\"odinger hierarchy
(\ref{eq:finBBGKY})
if $V$ is replaced with $V_1$, i.e.
\bey \label{eq:finBBGKYep}
\lefteqn{\gamma_{N,t}^{\e,(k)}(x_1, \ldots x_k;
    x_1', \ldots x_k')}  \\
    & = &  \cU_{N,\e, k}(t)\gamma_{N,0}^{(k)}
    (x_1, \ldots x_k;
    x_1', \ldots x_k')
 + (-i)\sum_{\ell=1}^k \int_0^t \rd s \; \cU_{N,\e,k}(t-s) \no \\
& &     \frac{N-k}{N}\Bigg[ \int \rd x_{k+1}
    \Big( V_1(x_\ell-x_{k+1})- V_1(x_\ell'-x_{k+1}) \Big) \no\\
   && \times \gamma_{N,s}^{\e,(k+1)}(x_1, \ldots x_k, x_{k+1};
    x_1', \ldots x_k', x_{k+1})\Bigg] \no
\eey
where $\cU_{N,\e,k}(t)\gamma
= e^{-itH_{N,\e}^{(k)}}\gamma e^{itH_{N,\e}^{(k)}}$
and
$$
    H_{N,\e}^{(k)} =- \frac{1}{2}\sum_{\ell=1}^k \Delta_\ell
    + \frac{1}{N} \sum_{\ell<j}^k V_1(x_\ell-x_j) \; .
$$

We now  consider the evolution of the cutoff dynamics
with smooth initial data.
We introduce the notation
$$
    L: = \frac{1}{N}\sum_{\ell=1}^N (I-\Delta_{x_\ell}) \;.
$$
We are ready to state the main propositions on the BBGKY hierarchy:

\begin{proposition}\label{thm:apriori} [Apriori bound]
Assume that the initial data of the $N$-body Schr\"odinger hierarchy
(\ref{eq:finBBGKYep})
\begin{equation}
    \Gamma_{N,0} : = \Big( \gamma_{N,0}^{(1)}, \gamma_{N,0}^{(2)},
    \ldots \gamma_{N,0}^{(N)}, 0, 0, \ldots \Big) \in \cC
\label{def:Gamma0}
\end{equation}
forms a consistent sequence of normalized  density matrices, i.e.
$\gamma_{N,0}^{(k)} = \Tr_{k+1} \; \gamma_{N,0}^{(k+1)}$,
$k=1, 2, \ldots, N-1$ and $\Tr \; \gamma_{N,0}^{(N)}=1$.  Assume that
\begin{equation}\label{highsbound}
 \Tr \;  L^k \gamma^{(N)}_{N,0}  \le \nu^k_0, \qquad k=1,2, \ldots N
\end{equation}
for some constant $\nu_0>0$ independent of $N$.
Let $\e > 0$ be any fixed constant and
\begin{equation}
    \Gamma_{N,t}^\e : = \Big( \gamma_{N, t}^{\e, (1)},
    \gamma_{N, t}^{\e, (2)},
    \ldots \Big)
\label{def:Gammat}
\end{equation}
be the solution to the $N$-body Schr\"odinger
hierarchy (\ref{eq:finBBGKYep}) with initial data $\Gamma_{N, 0} $
and Hamiltonian $H_{N, \e} $. Then there exists a $\nu>0$ depending
only on $\nu_0$,  and for any $k$ fixed there
is a constant $N(\e, k)$ such that
$\Gamma_{N,t}^\e$
satisfies the a priori estimate
\begin{equation}
 \| \gamma^{\e, (k)}_{N, t} \|_{\cH^{1,(k)}}\le \nu^k
\label{eq:apriori}
\end{equation}
for  $N \ge N(\e, k)$.
Therefore,
$\Gamma_{N,t}^\e$ forms a pre-compact
sequence in
$L^\infty(\bR_+,\cD)$
(with the w$^*$ topology) by Lemma \ref{lemma:wc}.
\end{proposition}

For  $\nu>0$ define the norm for $\Gamma = \big(\gamma^{(1)},
\gamma^{(2)},\ldots \big)\in \cD$
$$
\| \Gamma \|_\nu : =
\sum_{k=1}^\infty \nu^{-k} \| \gamma^{(k)}\|_{\cH^{1, (k)}} \; .
$$

\begin{proposition}\label{prop:conv} [Convergence to the infinite
hierarchy]
Assume the initial data (\ref{def:Gamma0})
satisfies the estimates \eqref{highsbound}.
Let $\e > 0$ be any fixed constant. Then any w$^*$ limit point
$\Gamma_{\infty,t}\in\cD$ of the sequence
$\{ \Gamma_{N,t}^\e \}_{N=1,2,\ldots}\subset \cD$ satisfies the infinite
BBGKY hierarchy \eqref{eq:BBGKY} and for some $\nu$ large enough
$\| \Gamma_{\infty, t}\|_\nu$
is uniformly bounded for all $t$.
\end{proposition}

At this stage we do not know that the limit point
$\Gamma_{\infty, t} = \Big( \gamma_{\infty,t}^{(1)}, \gamma_{\infty, t}^{(2)},
\ldots \Big)$ forms a consistent family of density matrices. This
will follow from the uniqueness of the solution to the nonlinear
Schr\"odinger equation (\ref{eq:NLS})
and the uniqueness result below.

\begin{proposition}\label{lemma:unique} [Uniqueness of the infinite
BBGKY hierarchy]
Let  $T>0$ be any fixed time. Then for any $\nu>0$
the infinite BBGKY hierarchy (\ref{eq:BBGKY})
has at most one solution in the set
$L^\infty\big ( [0,T],(\cD, \|\cdot\|_\nu)\big )$.

\end{proposition}

\bigskip

\subsection{Approximation by smooth initial data}

For any $\kappa>0$ let  $\psi_0^\kappa: = e^{\kappa \Delta}\psi_0$.
This smoothed initial data  satisfies the following bound
\begin{equation}
   \|\psi_0-\psi_0^\kappa\|  \leq C\kappa \|\psi_0\|_{H^2}
\label{eq:conv}
\end{equation}
using a simple estimate in Fourier space
$$
   \|\psi_0-\psi_0^\kappa\|^2 = \int (e^{-\kappa p^2}-1)^2 |\wh \psi_0(p)|^2
    \rd p \leq C\kappa^2 \int p^4|\wh \psi_0(p)|^2 \rd p \; .
$$

For $\delta>0$, let
$$
\Psi_{N,0}^\delta (x_1, \ldots x_N): = \prod_{j=1}^N \psi_0^{\delta/N}
 (x_j)
$$
denote the $\delta/N$ regularized initial wave function.
Let $\gamma_{N,0}^{\delta, (N)}$ be
the projection matrix onto $\Psi_{N,0}^\delta$.
The main reason to regularize the initial data
is the following bound, whose proof is postponed.

\begin{proposition}\label{prop:com}  There exists a  constant
$C$ such that for any $\delta >0$ and $k\ge 0$,
\begin{equation}
\Tr\Big[ L^k \gamma_{N, 0}^{\delta, (N)}\Big]
= \langle \Psi_{N,0}^\delta, L^k \Psi_{N,0}^\delta\rangle
\leq    \Big(C \|\psi_0\|_{H^2}^2\Big)^k
\label{eq:energypower}
\end{equation}
whenever $N$ is sufficiently large depending on $\delta$ and $k$.
\end{proposition}

The proof of this Proposition will be given in Sect \ref{sect:apriori}.
{F}rom now on we assume  that the one-particle
wave function of the initial data is  $\psi_0\in H^2(\bR^3)$.
We can now apply the previous setup of the $\e/\sqrt N$ cutoff dynamics
(Section \ref{sec:cut}) to the regularized initial data. Let
$$
    \Psi_{N,t}^{\delta} = e^{-it H_{N}} \Psi_{N,0}^\delta\; , \quad \mbox{and}
    \quad
    \Psi_{N,t}^{\delta,\e} : = e^{-it H_{N,\e}} \Psi_{N,0}^\delta
$$
be the solutions to the Schr\"odinger equation \eqref{N-Sch}
with the original Hamiltonian and with the regularized Hamiltonian,
$H_{N,\e}$, respectively.
 Denote the corresponding density matrices by
$\gamma^{\delta, (k)}_{N, t}$ and  $\gamma^{\delta,\e, (k)}_{N, t}$.
{F}rom Propositions \ref{thm:apriori}-\ref{prop:com} and from
the fact that $\Gamma_{\psi_t}$ (\ref{Gammapsi}) solves the infinite
BBGKY hierarchy (\ref{eq:BBGKY}), we have
the following corollary. Notice that the BBGKY hierarchy is independent
of the cutoff $\e/\sqrt N$
and the initial data for the BBGKY hierarchy is independent
of the smoothing parameter $\delta/N$. Both regulatizations
disappear after we have taken the limit $N \to \infty$.

\begin{corollary}\label{cor:main}
Suppose  that $\psi_0 \in H^2(\bR^3)$ and let $\psi_t$ be the solution
to (\ref{eq:NLS}). There exists a
constant $\nu$ depending on $\|\psi_0\|_{H^2}$
such that for any $k$ and $\delta>0$
\begin{equation}
    \| \gamma^{\delta, \e , (k)}_{N,t} \|_{\cH^{1, (k)}}
    \leq \nu^k
\label{eq:apr}
\end{equation}
for sufficiently large $N\ge N(\delta, \e, k)$. Furthermore,
$$
    \Gamma_{N,t}^{\delta,\e} \to \Gamma_{\psi_t}
$$
 as $N\to\infty$,
in the $w^*$-topology of  $L^\infty(\bR_+,\langle t \rangle^{-1} \rd t, \cD)$.
In other words, for each $k$
$$
    \gamma^{\delta, \e ,(k)}_{N, t} \to \gamma^{(k)}_{\psi_t}
$$
in the weak* topology of $L^\infty(\bR_+,\langle t \rangle^{-1} \rd t,
 \cH^{1}(H^{\otimes k}) )$ as $N\to\infty$.
\end{corollary}

We now control the deviation due to the cutoff and smoothing.
The  next Lemma shows that the $N$-body wave functions for the $\delta/N$
regularized  and the original initial data are close for all time.

\begin{lemma}\label{lemma:Nbodyapprox}
Suppose the initial data $\psi_0 \in H^2(\bR^3)$. Then we have
\begin{equation}
    \lim_{\delta\to0}\limsup_{N\to\infty}\sup_{t}
    \| \Psi_{N,t} - \Psi_{N,t}^\delta\| =0 \; .
\end{equation}
\end{lemma}

\noindent
{\it Proof.} By unitarity, it is sufficient to consider $t=0$;
$$
    \| \Psi_{N,0} - \Psi_{N,0}^\delta\| \leq
    \sum_{j=1}^N \| \psi_0\|^{j-1} \;
    \| \psi_0 - \psi_0^{\delta/N}\| \;
    \Big\| \psi_0^{\delta/N}\Big\|^{N-j-1}
    \leq \delta \|\psi_0\|_{H^2}
$$
using (\ref{eq:conv}) and $\|\psi_0^\kappa\|\leq \|\psi_0\|=1$.
\myendproof

Finally, the following lemma controls the difference between
the original dynamics and the $\e/\sqrt N$-regularized dynamics.

\begin{lemma}\label{lemma:cutoff} For any $\e$ fixed, we have
$$
    \sup_\delta \limsup_{N\to\infty}
    \|  \Psi_{N,t}^{\delta,\e} - \Psi_{N,t}^\delta\|^2 \leq
    C (\psi_0) \e t \; ,
$$
where $C(\psi_0)$ depends on the $H^2$-norm of $\psi_0$.
\end{lemma}

The proof of this Lemma will be given in Section \ref{sect:cutoff}.
Combining these Lemmas, we have
$$
\lim_{\e \to 0} \lim_{\delta\to 0}\limsup_{N\to\infty}\sup_{t}
   \langle t \rangle^{-1} \| \Psi_{N,t}^{\delta,\e} - \Psi_{N,t}\| =0\; .
$$
Thus we have for any $k\ge 1$
$$
\lim_{\e \to 0} \lim_{\delta\to 0}\limsup_{N\to\infty}
    \sup_t \langle t \rangle^{-1}
\| \gamma^{\delta,\e, (k)}_{N, t} -
\gamma^{(k)}_{N, t} \|_1 = 0\; .
$$
Since the trace norm is stronger than the $w^*$-topology in $\cL^1(H^{\otimes
(k)})$, Theorem \ref{thm:main}
follows from Corollary \ref{cor:main} and Lemmas \ref{lemma:Nbodyapprox}
and \ref{lemma:cutoff}.
 Notice that
 the limit in Corollary \ref{cor:main} was controlled in
 the $w^*$ Sobolev norm. The control in Lemmas \ref{lemma:Nbodyapprox}
 and \ref{lemma:cutoff} was in strong sense but without control
 on the derivatives. This is why the final result is only
 in the $w^*$-topology of trace class operators. \myendproof

\bigskip

\section{Apriori bounds}\label{sect:apriori}
\setcounter{equation}{0}

In this section we first prove Proposition  \ref{prop:com}
then  Proposition \ref{thm:apriori}. Recall that Proposition
\ref{prop:com} guarantees that the smoothed initial data
satisfies the apriori bound (\ref{highsbound}) for the initial
condition in  Proposition \ref{thm:apriori}.

\bigskip

\noindent
{\it Proof of Proposition \ref{prop:com}}:
Let $\bn= (n_1, n_2, \ldots n_\ell)$ be a sequence of positive integers,
 let $k(\bn) = n_1+ \ldots +n_\ell$ and $\ell(\bn)=\ell$.
 We also define $L_j: = \frac{1}{N}(I-\Delta_{x_j})$, clearly
 $L= \sum_{j=1}^N L_j$.
The following identity is easily checked
\begin{equation}
    L^k = \sum_{\ell=1}^k \sum_{\bn \; : \; k(\bn)=k\; : \; \ell(\bn) = \ell}
    \quad \sum_{i_1, \ldots i_\ell =1 \; : \; disjoint}^N
    L_{i_1}^{n_1} L_{i_2}^{n_2} \ldots L_{i_\ell}^{n_\ell}\; .
\label{eq:id}
\end{equation}
We have (recall  $\kappa = \delta/N$)
\bey
    \Big\langle \psi^\kappa_0, (I-\Delta)^{n}
    \psi_0^\kappa\Big\rangle &=&
    \int (1+ p^2)^n e^{-\kappa p^2} |\wh\psi_0(p)|^2 \rd p \no\\
    &\leq& 2^n \| \wh \psi_0\|^2
    + 2^n \int p^{2n} e^{-\kappa p^2} |\wh\psi_0(p)|^2 \rd p\no\\
    &\leq&    C^n (1 + n! \; \kappa^{2-n})
     \| \psi_0\|_{H^2}^2 \; .
\eey

Therefore
\bey
    \lefteqn{\sum_{i_1, \ldots i_\ell =1 \; : \; disjoint}^N
     \Big\langle  \Psi_{N,0}^{\delta},
    L_{i_1}^{n_1} L_{i_2}^{n_2} \ldots L_{i_\ell}^{n_\ell}
     \Psi_{N,0}^{\delta}\Big\rangle}  \\
    &=& \frac{N!}{(N-\ell)! N^{k}}\prod_{j=1}^\ell \Big
    \langle \psi^\kappa_0, (I-\Delta)^{n_j}
    \psi_0^\kappa\Big\rangle\no\\
    &\leq& N^{\ell-k} \Big( C \| \psi_0\|_{H^2}^2\Big)^k
     \prod_{j=1}^\ell (1 + n_j! \; \kappa^{2-n_j}) \no\\
    & = &\Big( C \| \psi_0\|_{H^2}^2\Big)^k
     \prod_{j=1}^\ell \Big(
    N^{1-n_j}  + n_j! \; N^{-1}\delta^{2-n_j}\Big)\no\\
    &\leq& \Big( C\| \psi_0\|_{H^2}^2\Big)^k \times \left\{
\begin{array}{lll}
1 & \mbox{if} & \ell = k \;\;\; (\mbox{i.e.} \; n_1=n_2=\ldots =1) \cr
k!\;\langle \delta^{-k}\rangle N^{-1}  & \mbox{if} & \ell < k
\end{array}
\right. \; . \no
\eey
Hence from this estimate, (\ref{eq:id}) and from the exponential
bound $C^k$ on number of possible $\bn$ sequences with $k(\bn)=k$ we obtain
\begin{equation}
     \Big\langle  \Psi_{N,0}^{\delta}, L^k
     \Psi_{N,0}^{\delta}\Big\rangle \leq
    \Big( C\| \psi_0\|_{H^2}^2\Big)^k
    \Big( 1+  k!\;\langle \delta^{-k}\rangle N^{-1}\Big)
\label{eq:initest}
\end{equation}
and (\ref{eq:energypower}) follows. This completes the proof of
Proposition \ref{prop:com}. \myendproof

\bigskip

\noindent
{\it Proof of Proposition \ref{thm:apriori}.}
We define
\begin{equation}
    Q_i = \frac{1}{N} (\beta^2 - \sfrac{1}{2}\Delta_{x_i})\; , \qquad
    Q:=\sum_{i=1}^N Q_i
\label{def:Q}
\end{equation}
where the constant $\beta\ge 2$ will be chosen
later. We also  introduce $U_{ij}: = N^{-2} V_1(x_i-x_j)$ (for
$i\neq j$), $U_{ii} =0$ and $U:= \sum_{1\leq i < j \leq N}
U_{ij}$. Thus
$$
    H_{N,\e} + \beta^2N = N(Q+U) \; .
$$
We also define
$$
    Q^{(k)}: = \sum_{i_1, i_2, \ldots i_k=1\; : \; disjoint}^N
        Q_{i_1} Q_{i_2} \ldots Q_{i_k} \; .
$$
Note that $Q, Q^{(k)}$ and $U$ depend on $N$, but this fact
is suppressed in
the notation.
The key Proposition is the following result to compare
$Q^k$ with $H_{N,\e}^k$:

\begin{proposition}\label{lemma:compare}
There exist a constant $\beta$ (depending only on $\mu$)
 such that for any  $k$ and any $N\ge N( \e,k)$
we have the following operator inequalities
\begin{equation}
    C_1^k Q^{k}\leq
      (Q+U)^k  \leq C_2^k  Q^k
\label{eq:compare}
\end{equation}
with some positive constants $C_1, C_2$.
\end{proposition}

First we show how this Proposition implies Proposition
\ref{thm:apriori}. The assumptions (\ref{highsbound}) and
(\ref{eq:compare}) imply that
$$
    \Tr \; \gamma_{N,0}^{(N)} (H_{N, \e}+\beta^2N)^k \leq
    C_3^k(\nu_0^k + \beta^{2k}) N^k
$$
for $k=1, 2, \ldots N$ if $N$ is large enough.
Since any function of the Hamiltonian is conserved along the time evolution,
we have
$$
    \Tr \; \gamma_{N,t}^{\e, (N)} (H_{N, \e}+\beta^2N)^k \leq
    C_3^k(\nu_0^k + \beta^{2k})     N^k \; .
$$
By (\ref{eq:compare}) and $Q^{(k)}\leq Q^k$
we obtain
$$
    \Tr \; \gamma_{N,t}^{\e, (N)} Q^{(k)} \leq
    C_4^k(\nu_0^k + \beta^{2k}) \; .
$$
Using the total symmetry of the $N$-body density matrix and $\beta\ge 1$,
$$
    \|\gamma_{N, t}^{\e, (k)} \|_{\cH^{1, (k)}}
    \leq  2^k \; \Tr \; \gamma_{N,t}^{\e, (N)} Q^{(k)}
$$
and (\ref{eq:apriori}) follows. The pre-compactness of the
sequence $\Gamma_{N, t}^\e$ follows from
the multivariable analog of Lemma \ref{lemma:wc}
and a diagonal selection procedure. \myendproof

\bigskip
\noindent
{\it Proof of Proposition \ref{lemma:compare}}:
We proceed by a step two induction on $k$.
 The statement
(\ref{eq:compare})  for $k=0$ is trivial.
For $k=1$  the bounds follow from
\begin{equation}
    |U|\leq C\beta^{-1}Q
\label{eq:potdom}
\end{equation}
which is a consequence of Hardy's inequality in three dimensions
(Uncertainty Principle Lemma in Vol.II. Section X.2. of
\cite{RS})
\begin{equation}
   \frac{1}{4|x|^2}\leq -\Delta_x\;,    \qquad x\in \bR^3\; .
\label{eq:hardy}
\end{equation}

%$|U_{ij}|\leq \frac{1}{2N\beta} (Q_i+Q_j)$ following from
%Hardy's inequality $|V_1|\leq \mu|x|^{-1} \leq C\mu(I-\Delta)$.

Assuming (\ref{eq:compare}) is valid for some $k$, we
show it for $k+2$. Writing
$(Q+U)^{k+2} = (Q+U) (Q+U)^{k} (Q+U)$ we have
$$
    C_1^k (Q+U) Q^{k} (Q+U)
    \leq (Q+U)^{k+2}\leq
     C_2^k (Q+U) Q^k (Q+U)\; .
$$
In order to compare $UQ^kU$ with $Q^{k+2}$ we will prove
the following lemma:

\begin{lemma}\label{lemma:UPU}
For $0\leq k\leq N/2$, $0\leq a <3$
\begin{equation}
    UQ^kU\leq \Bigg( C\beta^{-2} + C(k)\Big[ \beta^{2k}N^{-k} +
  N^{1-a/2}\e^{-2k-2+a}+ N^{-1}\Big]\Bigg) Q^{k+2}\; .
\label{eq:UQU}
\end{equation}
with constants depending on $a$ and we can set $C(0)=0$.
\end{lemma}

\noindent
{\it Remark:} A similar
 estimate holds for potentials much more singular than the Coulomb
potential. With essentially the same proof, one can establish it for
$V(x) \sim |x|^{-2+ \eta}$ for any $\eta > 0$.

\bigskip

{F}rom the Schwarz inequality $A^* B+ B^* A \le  A^* A + B^* B$,
with $A^*= \sqrt{2}U Q^{k/2}$, and $ B=-\sfrac{1}{\sqrt{2}}Q^{k/2+1}$,
 we have
$$
-\Big( U Q^{k}Q + Q Q^{k}U\Big)  \le 2 U Q^{k}U +\frac{1}{2} Q Q^{k}Q\; .
$$
Together with  Lemma \ref{lemma:UPU} with some
$2<a<3$ and $\beta$ large enough,
the lower bound in (\ref{eq:compare}) follows from
$$
     (Q+U) Q^{k} (Q+U)
    \ge  \frac{1}{2}Q Q^{k}Q - U Q^{k}U
    \ge \frac{1}{4}  Q^{k+2}
$$
as $N\ge N(\e, k)$. So $C_1=1/2$.

\bigskip

For the upper bound, we again let $2< a <3$. Similar arguments lead to
$$
    (Q+U) Q^k (Q+U)  \leq 2Q^{k+2} + 2 UQ^kU
     \leq C  Q^{k+2}
$$
for $N$ large enough,
which completes the proof of Lemma \ref{lemma:compare} \myendproof

\bigskip
\noindent
{\it Proof of Lemma \ref{lemma:UPU}}.
{F}rom Schwarz inequality, we obtain
\begin{equation}
    UQ^{k}U \leq
     N^2 \sum_{i,j=1}^N |U_{ij}| \;  Q^{k}
    \;  |U_{ij}|\; .
\label{eq:UQU1}
\end{equation}
For $k=0$,  (\ref{eq:UQU}) follows from
 \begin{equation}
    |U_{ij}|^2\leq C N^{-3} Q_j\leq C\beta^{-2}N^{-2}Q_iQ_j \; ,
\label{eq:noder}
\end{equation}
which is a consequence of Hardy's inequality  (\ref{eq:hardy}).

For $k\ge 1$ and for each fixed $i,j$ we write $Q= Q_i+Q_j + Q_{\wh{ij}}$.
By the weighted Minkowski inequality, we have for some constant
$C$
\begin{equation}
    Q^k \leq (C k)^k (Q_i^k + Q_j^k) + 2 Q_{\wh{ij}}^k \; .
\label{eq:mink}
\end{equation}

Let $D_i : = \sqrt{-1} N^{-1/2} \nabla_{x_i}$.
We have from Leibniz rule and Schwarz inequality
that there is a constant $C(k)$ such that
\begin{equation}
    |U_{ij}| \; D_i^{2k} \; |U_{ij}|
    \leq C D_i^k \;  |U_{ij}|^2 D_i^k  +
    C(k) \sum_{m=1}^k   D_i^{k-m} \; \Big| (D_i^{m}
     |U_{ij}|)\Big|^2  D_i^{k-m}\; .
\label{eq:Uijest}
\end{equation}
To estimate the derivative of $U_{ij}$ we notice that
\begin{equation}
    |\nabla^m_xV_1(x-y)|^2\leq C^m \Big( \sqrt{N}/\e \Big)^{2m+2-a}|x-y|^{-a}
\label{eq:dee}
\end{equation}
for $a\leq 2m+2$. We will choose $2<a<3$.
The singularity of $|x-y|^{-a}$ is controlled by
the following lemma whose proof is postponed:

\begin{lemma}\label{lemma:H11}
Let $W\in L^1 (\bR^3)$ be any nonnegative potential, then
\begin{equation}
    W(x-y) \leq C \| W \|_{L^1} (I-\Delta_x)(I-\Delta_y)
\label{eq:genW}
\end{equation}
on $L^2(\bR^3\times\bR^3)$.
\end{lemma}
Combining this lemma with (\ref{eq:dee}) we obtain $(m\ge 1)$
\begin{equation}
     \Big| (D_i^{m}  |U_{ij}|)\Big|^2 \leq C^mN^{-1-a/2}
    \e^{-2m-2+a}Q_iQ_j \; .
\label{eq:dm}
\end{equation}

We have $Q_i^k \leq C(k)\big[  D_i^{2k} +(\beta^2N^{-1})^k\big]$.
Hence from (\ref{eq:noder}),  (\ref{eq:Uijest}),
and (\ref{eq:dm})  we obtain
$$
    |U_{ij}| \; Q_i^k \; |U_{ij}|
    \leq C(k)N^{-2}
    \Big(  (\beta^2N^{-1})^k
    +  N^{1-a/2}\e^{-2k-2+a}+ N^{-1}\Big)Q_j(Q_i^{k}+ Q_i)\; .
$$
Using $Q= \sum_j Q_j$, $\sum_i Q_i^k \leq Q^k$ and $Q\ge I$,
\begin{equation}
     N^2 \sum_{i,j=1}^N |U_{ij}| \;  Q_i^{k}
    \;  |U_{ij}| \leq
    C(k)\Big(  (\beta^2N^{-1})^k +
     N^{1-a/2}\e^{-2k-2+a}+ N^{-1}\Big)Q^{k+1} \; .
\label{eq:Qi}
\end{equation}
Finally, we use that $U_{ij}$ and $ Q_{\wh{ij}}$ commute
and the estimate (\ref{eq:noder}) to bound
\begin{equation}
     N^2 \sum_{i,j=1}^N |U_{ij}| \;  Q_{\wh{ij}}^k
    \;  |U_{ij}| \leq  N^2\sum_{i,j=1}^N C\beta^{-2}N^{-2}
    Q_iQ_j Q_{\wh{ij}}^k \leq C\beta^{-2} Q^{k+2} \; .
\label{eq:Qall}
\end{equation}
Hence from (\ref{eq:mink}), (\ref{eq:Qi}),  (\ref{eq:Qall})  and $Q\ge I$
the statement (\ref{eq:UQU}) follows.
This finished the proof of Lemma \ref{lemma:UPU} \myendproof

\bigskip

\noindent
{\it Proof of  Lemma \ref{lemma:H11}.} Let
\begin{equation}
    H(x): = \int_{\bR^3} \frac{e^{ipx}}{\sqrt{1+p^2}} \; \rd p \; ,
\label{Hdef}
\end{equation}
then $H(x) \sim |x|^{-2}$ around the origin, smooth outside
 and it decays faster than any polynomial at infinity.
Clearly (\ref{eq:genW}) is equivalent to
\begin{equation}
    \frac{1}{\sqrt{ (I-\Delta_x)(I-\Delta_y)}} W(x-y)
    \frac{1}{\sqrt{ (I-\Delta_x)(I-\Delta_y)}} \leq C \| W \|_{L^1} \; .
\label{eq:W}
\end{equation}

Let $f\in L^2(\bR^3\times\bR^3)$, then
\bey
\lefteqn{\Bigg\langle f, \frac{1}{\sqrt{ (I-\Delta_x)(I-\Delta_y)}} W(x-y)
    \frac{1}{\sqrt{ (I-\Delta_x)(I-\Delta_y)}} f \Bigg\rangle} \\
    &=& \int  \rd x \rd x' \rd x''
    \rd y \rd y' \rd y''\overline{f(x,y)} H(x-x')H(y-y') W(x'-y') \no\\
   &&\times H(x'-x'')H(y'-y'') f(x'', y'') \no \\
   &\leq& 2 \int \rd x \rd x' \rd x''
    \rd y \rd y' \rd y'' |f(x,y)|^2 \Bigg[ \frac{|x-x'|}{|x'-x''|}
     \frac{|y-y'|}{|y'-y''|}\Bigg]^{3/4}\no \\
  &&\times  \Big| H(x-x')H(y-y') W(x'-y')
    H(x'-x'')H(y'-y'')\Big|\no
\eey
by symmetry of $(x,y)$ and $(x'', y'')$ and a Schwarz inequality.
By the properties of $H$
$$
    \int \frac{|H(u)|}{|u|^{3/4}} \rd u \leq C
$$
hence $x'', y''$ integrals can be performed. We also define
 $G(u): = |u|^{3/4}|H(u)|$ and notice that $\| G \|_{L^2} \leq C$.
Finally, by Young's
inequality:
$$
    \sup_{x,y} \int G(x-x') |W(x'-y')| G(y'-y) \rd x' \rd y'
    \leq \| G \|_{L^2}^2 \| W\|_{L^1}
$$
we obtain (\ref{eq:W}). \myendproof

\bigskip

\section{Proof of  the convergence}
\setcounter{equation}{0}

In this section we prove Proposition \ref{prop:conv}.
The uniform boundedness of $\|\Gamma_{\infty, t}\|_\nu$
follows from (\ref{eq:apriori}) and  Lemma \ref{lemma:wc}.
To check that the limit satisfies (\ref{eq:BBGKY}),
it is sufficient test the equation
against trace class operators  $\cO \in\cL^1(H^{\otimes k})$
 on a fixed time interval $[0, T]$.
We note that $\Gamma_{N, t}^\e$ satisfies the
finite Schr\"odinger hierarchy with the cutoff potential
(\ref{eq:finBBGKYep}).

Since $\cO$
is compact, we have
$$
  \lim_{N \to \infty}   \Tr \;\cO \Big( \gamma_{N,t}^{\e,(k)}
    -\gamma_{\infty,t}^{(k)}\Big)
    = 0\; .
$$
This shows the convergence of the left  side of
(\ref{eq:finBBGKYep})  to that of (\ref{eq:BBGKY}).

For the convergence of the first term on the right side
of (\ref{eq:finBBGKYep}),
we now show that  $\cU_{N,k,\e}(t)$ converges to $\cU_k(t)$.
By one-step Duhamel expansion we have
\begin{equation}
     e^{isH_{N,\e}^{(k)}} -
    \Big(\prod_{j=1}^k e^{is\Delta_j}\Big)
    =   \frac{-i}{N} \sum_{\ell<j}^k \int_0^s
    e^{i(s-u)H_{N,\e}^{(k)}} V_1(x_\ell-x_j)
    \Big(\prod_{j=1}^k e^{iu\Delta_j}\Big)\; \rd u \;.
\label{eq:duh}
\end{equation}
{F}rom the triangle inequality, the unitarity of
$e^{i(s-u)H_{N,\e}^{(k)}}$ and
$|V_1|\leq C\e^{-1}N^{1/2}$, we can bound the
operator norm of the left side by
\bey\label{eq:evolu}
  \lefteqn{   \Bigg\| e^{isH_{N,\e}^{(k)}} -
    \Big(\prod_{j=1}^k e^{is\Delta_j}\Big)\Bigg\|} \qquad\qquad\qquad\no\\
    &\leq&   \frac{1}{N} \sum_{\ell<j}^k \int_0^s
    \Bigg\| e^{i(s-u)H_{N,\e}^{(k)}} V_1(x_\ell-x_j)
    \Big(\prod_{j=1}^k e^{iu\Delta_j}\Big)\Bigg\|\; \rd u \no\\
    &\leq& \frac{C k^2 s}{\e\sqrt{N}} \;.
\eey
Using  (\ref{eq:evolu}) the convergence of
$\gamma_{N,0}^{(k)}$, the fact that
$\Tr \gamma_{\infty,0}^{(k)} \leq
\Tr \gamma_{N,0}^{(k)} =1$ and  $\| \cO\|<\infty$, we thus have
$$
    \lim_{N\to\infty}
    \Bigg|  \; \Tr \; \cO  \Big[\cU_{N,k,\e}(t)\gamma_{N,0}^{(k)}
    -  \cU_k(t)
    \gamma_{\infty,0}^{(k)} \Big]\; \Bigg| =0 \; .
$$

\bigskip

We now show  the convergence of the second term on the right hand side
of (\ref{eq:finBBGKYep}). The prefactor $(N-k)/N$
can be replaced by 1 in the $N\to\infty$ limit, since
$$
    \Bigg| \frac{k}{N}  \int_0^t \! \Tr \; \cO  \cU_{N,k,\e}(t-s)
  \Bigg[ \int \rd x_{k+1}
    \Big( V_1(x_\ell-x_{k+1}) -V_1(x_\ell'-x_{k+1})\Big)
     \gamma_{N,s}^{\e, (k+1)}\Bigg] \rd s \Bigg|
$$
$$
    \leq \frac{2kt}{N} \| \cO \| \| V_1\|_\infty
    \sup_{s\in [0,t]}\Tr \,\, \gamma_{N,s}^{\e, (k+1)}
    = O(N^{-1/2}) \to 0
$$
for any fixed $k$ and $\e$ using that $\| V_1\|_\infty = O(N^{1/2})$.
We then show that
$$
  \lim_{N \to \infty}  \int_0^t \Tr \; \cO  \Bigg(\cU_{N,k,\e}(t-s)
  \Bigg[ \int \rd x_{k+1}
    \Big( V_1(x_\ell-x_{k+1}) -V_1(x_\ell'-x_{k+1})\Big)
     \gamma_{N,s}^{\e, (k+1)}\Bigg]
$$
$$
  -  \cU_k(t-s)
    \Bigg[ \int \rd x_{k+1} \Big( V(x_\ell-x_{k+1}) -V(x_\ell'-x_{k+1})\Big)
     \gamma_{\infty,s}^{(k+1)}
    \Bigg]\Bigg) \rd s = 0
$$
for any  $\ell\leq k$. Decompose this difference into three terms
$(I.) + (II.) + (III.)$ where
\begin{eqnarray}
  (I.) & = &  \int_0^t \Tr \;
     \Big(\cO \cU_{N,k,\e}(t-s)   - \cO \cU_k(t-s)  \Big)  \\
     &&  \times
    \Bigg[ \int \rd x_{k+1}\Big( V_1(x_\ell-x_{k+1}) -V_1(x_\ell'-x_{k+1})\Big)
     \gamma_{N,s}^{\e,(k+1)}\Bigg]\; \rd s \; ,\label{eq:I} \no \\
   (II.) & =  &  \int_0^t  \Tr \; \cO \cU_k(t-s)
    \Bigg[ \int \rd x_{k+1}\Big( V(x_\ell-x_{k+1}) -V(x_\ell'-x_{k+1})\Big)
    \no\\
     &&   \times
     \Big(\gamma_{N,s}^{\e,(k+1)} -
    \gamma_{\infty,s}^{(k+1)}\Big)\Bigg] \; \rd s \; , \label{eq:II} \\
   (III.) & = &   - \int_0^t  \Tr \; \cO  \cU_k(t-s) \label{eq:III} \\
   &&\times
     \Bigg[ \int \rd x_{k+1}\Big( V_2(x_\ell-x_{k+1}) -V_2(x_\ell'-x_{k+1})\Big)
     \gamma_{N,s}^{\e,(k+1)}\Bigg] \; \rd s \no\; .
\end{eqnarray}
We estimate these three terms separately.

\bigskip
\noindent
The first term is bounded by
\begin{eqnarray*}
    |(I.)| & \leq &  t \sup_{s\leq t}
    \Big\| \cO \cU_{N,k,\e}(t-s) - \cO \cU_k(t-s)\Big\| \;  \\
     &&   \times
    \sup_{s\leq t} \Tr \; \Big| \int \rd x_{k+1}
    \Big( V_1(x_\ell-x_{k+1}) -V_1(x_\ell'-x_{k+1})\Big)
     \gamma_{N,s}^{\e,(k+1)}\Big| \; .
\end{eqnarray*}
The norm $ \big\| \cO \cU_{N,k,\e}(t-s)   - \cO \cU_k(t-s) \big\| $
is bounded by
\bey
 \big\| \cO \cU_{N,k,\e}(t-s)- \cO \cU_k(t-s)  \big\|
 &\le&      \| \cO \|\; \sup_{s\leq t}
    \Big\| e^{isH_{N,\e}^{(k)}} -
    \Big(\prod_{j=1}^k e^{is\Delta_j}\Big)\Big\|  \no\\
&\leq& \frac{ C k^2  \langle t \rangle^2}{\e \sqrt{N}}      \| \cO \| \no
\eey
where we have used (\ref{eq:evolu}) in the last inequality.
To bound the trace term, from the triangle inequality we reduce
it to  estimate
$$
 \Tr \; \Big| \int \rd x_{k+1}  V_1(x_\ell-x_{k+1})
     \gamma_{N,s}^{\e,(k+1)}\Big|
$$
for $s\leq t$.
 Recall that we interpret the $x_{k+1}$ integration as a partial trace
over this variable. We can now use (\ref{eq:fubini}) and bound
this term by
$$
 \Tr \; \big|   V_1(x_\ell-x_{k+1})
     \gamma_{N,s}^{\e,(k+1)}\big|
$$
where the trace is now over all $k+1$ variables.
We define
$$
    S^{(k)}: = S_{x_1} S_{x_2}\ldots S_{x_k}\; .
$$
{F}rom  the inequality (\ref{eq:hardy}) we know that
\begin{equation}
    |V_1(x_\ell - x_{k+1})|^2 \leq |V(x_\ell - x_{k+1})|^2
    \leq C S_{x_{k+1}}^{2}
     \leq C [S^{(k+1)}]^2\;.
\label{eq:v1hardy}
\end{equation}
Hence
for any self-adjoint $\gamma \in \cH^{1, (k+1)}$ we have
\begin{equation}
    \Tr \Big| V_1(x_\ell-x_{k+1})
    \gamma\Big| = \Tr \sqrt{ \gamma V_1^2 \gamma}
    \leq  C\Tr \sqrt{ \gamma [S^{(k+1)}]^2 \gamma }
    =   C\Tr \sqrt{ S^{(k+1)}\gamma^2 S^{(k+1)}  }
\label{eq:v2e}
\end{equation}
using the cyclicity for the trace of the square
root (\ref{rootcycle}).
Since  the square root is monotonic in operator sense and $S^{(k+1)}\ge I$,
we obtain
\begin{equation}
 C\Tr \sqrt{ S^{(k+1)}\gamma^2 S^{(k+1)}  }
    \leq  C\Tr \sqrt{ S^{(k+1)}\gamma[S^{(k+1)}]^2\gamma S^{(k+1)}  }=
    C\| \gamma \|_{\cH^{1,(k+1)}} \; ,
\label{eq:v2est}
\end{equation}
where the last equality is the definition of the $H^1$ norm.
Therefore we showed that
$$
\Tr \; \Big| \int \rd x_{k+1}  V_1(x_\ell-x_{k+1})
     \gamma_{N,s}^{\e,(k+1)}\Big|
     \le C\| \gamma_{N,s}^{\e,(k+1)} \|_{\cH^{1, (k+1)} }
$$
which is uniformly bounded by (\ref{eq:apriori}),
hence the first term (I.) vanishes as $N\to\infty$.

\bigskip

The first half of the second term $(II.)$ is written as
$$
     \Tr \; \int_0^t\cO \cU_k(t-s) V(x_\ell-x_{k+1})
     \Big(\gamma_{N,s}^{\e,(k+1)} -
    \gamma_{\infty,s}^{(k+1)}\Big) \rd s\; ,
$$
where the trace refers to all $k+1$ variables.
In order to check that $(II.)$ vanishes in the limit $N \to \infty$,
it suffices to show that
$A(s): = [S^{(k+1)}]^{-1}  \cO \cU_k(t-s)  V_{\ell, k+1}
[S^{(k+1)}]^{-1}$
is in the space $L^\infty([0,t], \cK^{(k+1)})$, i.e. that it
is compact and is uniformly bounded for all $0\leq s \le t$.
Here $ V_{\ell, k+1}$ is the multiplication operator
by $V(x_\ell-x_{k+1})$.

For uniform boundedness, we only have to show that
$\cU_k(t-s)  V_{\ell, k+1}S_{k+1}^{-1}$ is uniformly bounded,
since $\cO$ is bounded. Here $S_{k+1}:= S_{x_{k+1}}$ for brevity.  From the
unitarity of
$\cU_k(t-s)$, it suffices  to prove that
\begin{equation}
\Big\|  V_{\ell, k+1}
   S_{k+1}^{-1} \Big\|
    \leq C \; .
\label{boundd}
\end{equation}
This follows from the  inequality \eqref{eq:v1hardy}.

For the compactness, it is sufficient to show that
$ S_{k+1}^{-1}\cO  \cU_k(t-s) V_{\ell, k+1} S_{k+1}^{-1}$
is actually Hilbert-Schmidt for all $s$.
 Recall that $\cO$ is  a
compact operator on $H^{\otimes k}$,
 i.e., in the space of the first $k$ variables,
and the unitary map
$ \cU_k(t-s)$ acts trivially on the  $(k+1)$-th variable.
Hence we can define $\cO_s : = \cO  \cU_k(t-s)$, which is a
Hilbert-Schmidt operator on $H^{\otimes k}$.
To emphasize that $\cO  \cU_k(t-s)$ is viewed as an operator
on $H^{\otimes (k+1)}$, we will write  $\cO_s \otimes I_{k+1}$.
The identity $I_{k+1} $ acts
on the $(k+1)$-th variable.

We introduce $X:=(x_1, \ldots x_k)$ and $Y=(y_1, \ldots y_k)$,  and
we compute the Hilbert-Schmidt norm of
$ S_{k+1}^{-1}\cO  \cU_k(t-s) V_{\ell, k+1} S_{k+1}^{-1}$:
\bey
\lefteqn{\Tr (\cO_s \otimes I_{k+1} ) S_{k+1}^{-1} V_{\ell, k+1} S_{k+1}^{-2}
     V_{\ell, k+1} S_{k+1}^{-1}(\cO_s \otimes I_{k+1} )} \qquad
    \qquad\qquad \no\\
    &=&\Tr (\cO_s \otimes I_{k+1} )  V_{\ell, k+1} S_{k+1}^{-2}
     V_{\ell, k+1} S_{k+1}^{-2}(\cO_s \otimes I_{k+1} ) \qquad\qquad \no\\
    &=& \!\!\! \int \! \cO_s (X; Y)
     V(y_\ell - x_{k+1}) \frac{1}{I-\Delta}(x_{k+1} - y_{k+1})
    V(y_{k+1} - y_\ell) \no \\
    && \times
     \frac{1}{I-\Delta}
    (y_{k+1}- x_{k+1})\cO_s (Y; X) \rd X \rd Y \rd x_{k+1} \rd y_{k+1}\no
    \\
    &\leq & C  \int |\cO_s (X; Y) |^2
     |V(y_\ell - x_{k+1})|^2 \no\\
    &&\times \Bigg|
     \frac{1}{I-\Delta}(y_{k+1}-  x_{k+1})\Bigg|^2
     \rd X \rd Y \rd x_{k+1} \rd y_{k+1} \no \\
    &\leq & C
      \int |\cO_s (X; Y) |^2 \rd X \rd Y < \infty\; ,
\eey
since $\cO_s$ is Hilbert-Schmidt. We used a Schwarz inequality
in the third line and the fact that
$$
    \int
     \Big| \frac{1}{I - \Delta}
    (z)\Big|^2 \rd z = \int \frac{\rd p}{(1+p^2)^2} <\infty \; .
$$
where $(I-\Delta)^{-1}(x-y)$ denotes the translation invariant
kernel of the operator  $(I-\Delta)^{-1}$.
This completes the estimate of the second term $(II.)$.

\bigskip

Finally the third term is estimated as
$$
    (III.)\leq 2t\|\cO\| \sup_{s\leq t}
    \; \Tr \; \Big| \int \rd x_{k+1}V_2(x_1-x_{k+1})
     \gamma_{N,s}^{\e, (k+1)}\Big| \; .
$$
{F}rom (\ref{eq:fubini}) and Lemma \ref{lemma:H11}
we can bound the trace term by
$$
    \Tr  \Big| V_2 \gamma\Big|
    = \Tr \sqrt{ \gamma V_2^2 \gamma }
    \leq C\e^{1/2} N^{-1/4} \Tr \sqrt{ \gamma S_1^2 S_{k+1}^2
     \gamma }
$$
where we introduced $\gamma: = \gamma_{N,s}^{\e, (k+1)}$
for brevity and we computed $\| V_2^2 \|_{L^1} \leq  C\e N^{-1/2}$.
 Clearly
$$
    \Tr \sqrt{ \gamma S_1^2 S_{k+1}^2
     \gamma } \leq \Tr \sqrt{ \gamma [S^{(k+1)}]^2
     \gamma }
    \leq  C\| \gamma \|_{\cH^{1,(k+1)}} \;
$$
as in (\ref{eq:v2e})-(\ref{eq:v2est}), which is uniformly
bounded in $N$ by (\ref{eq:apriori}).
Therefore $(III.)$ goes to zero as $N\to \infty$,
and we have proved  Proposition \ref{prop:conv}.
\myendproof.

\section{Uniqueness of the infinite hierarchy}
\setcounter{equation}{0}

\noindent
{\it Proof of Lemma \ref{lemma:unique}.}
Since the equation is linear,
it is sufficient to show that the  solution $\Gamma_t$
up to time $T$ is identically zero
if the initial condition is $\Gamma_0=0$
and if for some $K$ the apriori bound $\|\Gamma_t \|_\nu \leq K$
holds for $t\leq T$.
Notice that it is sufficient to show that uniqueness holds for
a short time, $t\le T(\nu)$, then one can extend
it up to time $T$ since the apriori bound
holds uniformly for $t\leq T$.

We need the following lemma whose proof is postponed.

\begin{lemma}\label{lemma:hardy}
 For any $\gamma\in \cH^{1, (k+1)}$, $k\ge 1$ and $\ell\leq k$
$$
    \Tr_{X} \; \Big| S^{(k)} \int \rd y \; V(x_\ell-y) \gamma(X, y ; X',y)
     S^{(k)}\Big|\leq
  C\| \gamma \|_{\cH^{1,(k+1)}} \; .
$$
where $X=(x_1, \ldots x_{k})$ for brevity and recall  $S^{(k)}  = S_{x_1}
S_{x_2}\ldots S_{x_k}$.
\end{lemma}

Armed with this estimate, we obtain from (\ref{eq:BBGKY}) and
$\gamma_0^{(k)}=0$ that
\bey
    \big\| \gamma^{(k)}_t \big\|_{\cH^{1, (k)}}
    & = &\Tr \; \Big| S^{(k)}\gamma^{(k)}_t S^{(k)}\Big| \\
   &\leq&
    2\sum_{\ell=1}^k \int_0^t \rd s \;
    \Tr \; \Big| S^{(k)} \Tr_{x_{k+1}}\big[ V(x_\ell-x_{k+1})
    \gamma^{(k+1)}_s\big]  S^{(k)}\Big| \no\\
   &\leq & C k \int_0^t \rd s
    \| \gamma^{(k+1)}_s \|_{\cH^{1, (k+1)}} \; . \no
\eey
Let $A(k,t): = \big\| \gamma^{(k)}_t \big\|_{\cH^{1, (k)}}$.
After iteration, we have
\begin{equation}
    A(k,t)
    \leq C^n k(k+1) \ldots (k+n-1)
    \int_0^t \rd s_1 \int_0^{s_1} \rd s_2 \ldots
    \int_0^{s_{n-1}} \rd s_{n}\;  A(n+k, s_n)\;.
\label{eq:iter}
\end{equation}
Since $\Gamma_t$ is bounded in $(\cD, \|\cdot \|_\nu)$, there exists
some $K$ such that
$$
    A(n,s) \leq K\nu^n
$$
for all $0\le s \le T$. After using this estimate in  (\ref{eq:iter}),
the multiple
time integration is bounded by $t^n/n!$. This $n!$ compensates
the combinatorial prefactor $k(k+1) \ldots (k+n-1)$ and the result is
$$
    A(k,t) \leq K(2\nu)^{n+k} (Ct)^n  \leq  K(2\nu)^k (C\nu t)^n
$$
for any $n$. If $t < T(\nu):= (C\nu)^{-1}$, then letting $n\to\infty$
we obtain that $A(k,t) =0$.
 \myendproof

\bigskip

\noindent
{\it Proof of Lemma \ref{lemma:hardy}.} Without loss of generality
 we can assume that $\ell=1$ and we set $x=x_1$,
i.e. $X=(x, x_2, \ldots x_k)$.
We define  $\wh S : = S_{x_2} S_{x_3} \ldots S_{x_k}$
and let $\wh\gamma: = \wh S\gamma \wh S$.
Let $\wt V$ be the multiplication operator by the function
$(x,y)\mapsto V(x-y)$ on $L^2(\bR^{3k}_X \times\bR^3_y)$
and let $P_x = (P_{x,1}, P_{x, 2}, P_{x, 3})$ be the components of momentum
operator in the $x$ variable, i.e., $P_{x, j}:= -i\nabla_j$.
$P_y$ is similarly defined.

We denote $R:=\Tr_y  (\wt V\wh \gamma)$ the partial trace
of $\wt V\wh\gamma$,
 then we have to estimate
$$
     \Tr_{X} \; \Big| S^{(k)} \int \rd y \;  V(x_1-y) \gamma(X, y ; X',y)
     S^{(k)}\Big| = \Tr_{X}  \; \Big| S_x R S_x\Big| \; .
$$

Using  the identity $S_x^2= I +\sum_{j=1}^3 P_{x, j}^2$
and Lemma \ref{lemma:traces} from the Appendix,  we  estimate
\bey\label{PS}
     \Tr_X \; \Big| S_x R S_x\Big|
     &=& \Tr_X \Big[ S_x R S_x^2 R S_x
     \Big]^{1/2} \\
    & \leq& C \Tr_X \Big[ S_x R^2 S_x
     \Big]^{1/2} + C \sum_{j=1}^3
      \Tr_X \Big[ S_x R P_{x, j}^2
      R S_x
     \Big]^{1/2} \no \\
      & =&  C \Tr_X | RS_x |
    + C \sum_{j=1}^3
      \Tr_X \Big|  P_{x, j} R S_x \Big| \no\\
    &=& C \Tr_X \Big|\Tr_y  (\wt V\wh\gamma) S_x \Big|
    + C \sum_{j=1}^3
      \Tr_X \Big|  \Tr_y P_{x, j}\wt V\wh\gamma S_x \Big| \; . \no
\eey
Hardy's inequality (\ref{eq:hardy}) implies that
\begin{equation}
    \wt V^2\leq S_x^2 \;, \qquad \wt V^2\leq S_y^2 \;.
\label{eq:Whardy}
\end{equation}
In the first term of (\ref{PS})
we use (\ref{eq:fubini}), (\ref{eq:Whardy})
and the fact that $S_x$ and $S_y$ commute
to obtain
\bey
    \Tr_X \Big|\Tr_y  (\wt V\wh\gamma) S_x \Big| &\leq&
    \Tr \Big|\wt V\wh\gamma S_x \Big|
    =\Tr \Big(S_x\wh\gamma \wt V^2\wh\gamma S_x \Big)^{1/2} \no\\
    &\leq&\Tr\Big(S_x\wh\gamma S_x ^2\wh\gamma S_x \Big)^{1/2} \no\\
    &\leq& \Tr \Big(S_x\wh\gamma S_x ^2S_y^2\wh\gamma S_x \Big)^{1/2}
    =\Tr \Big(S_xS_y\wh\gamma S_x ^2\wh\gamma S_xS_y \Big)^{1/2}\no\\
    &\leq&\Tr\Big(S_xS_y\wh\gamma S_x ^2S_y^2\wh\gamma S_xS_y \Big)^{1/2}
    = \Tr \big| S_xS_y \wh\gamma S_yS_x \big| \no\\
   &=& \| \gamma \|_{\cH^{1, (k+1)}} \no
\eey
where $\Tr$ denotes trace in all variables $X, y$.

For the second term in (\ref{PS}) we first
notice the following identity
$$
    P_{x, j} \wt V = -P_{y,j} \wt V + \wt V P_{x, j} + \wt V P_{y,j}
$$
Using $\Tr|A+B|\leq \Tr|A|+ \Tr|B|$, we obtain for
 each $j=1,2,3$
\bey
    \Tr_X \Big|  \Tr_y P_{x, j}\wt V\wh\gamma S_x \Big|
    &\leq&\Tr_X \Big|  \Tr_y P_{y, j}\wt V\wh\gamma S_x \Big|\label{eq:threeterm} \\
   && + \Tr_X \Big| \Tr_y \wt VP_{x, j}\wh\gamma S_x \Big|
    +\Tr_X\Big| \Tr_y
        \wt VP_{y, j}\wh\gamma S_x \Big|\no\\
  &\leq&\Tr \; \Big| \wt V\wh\gamma S_xP_{y,j} \Big|
    + \Tr\; \Big| \wt VP_{x, j}\wh\gamma S_x \Big|
    +\Tr\; \Big|
        \wt VP_{y, j}\wh\gamma S_x \Big| \no \; .
\eey
Notice that in the first term we  used the
cyclicity of the partial trace (\ref{partcycle})
from the Appendix
before applying (\ref{eq:fubini}).

The estimate of the three terms in (\ref{eq:threeterm}) are similar
by using the inequalities (\ref{eq:Whardy}) appropriately.
The estimate of the first term is
\bey
    \Tr \Big| \wt V\wh\gamma S_xP_{y,j} \Big|
    &=& \Tr\Big( P_{y,j} S_x \wh\gamma \wt V^2\wh\gamma S_xP_{y,j}
     \Big)^{1/2}\no\\
   & \leq&\Tr \Big( P_{y,j} S_x \wh\gamma S_x^2 S_y^2
    \wh\gamma S_xP_{y,j} \Big)^{1/2} \no\\
   & =&\Tr \Big( S_xS_y  \wh\gamma S_x^2 P_{y,j}^2
    \wh\gamma S_x S_y  \Big)^{1/2} \no\\
   & \leq& \Tr \Big( S_xS_y  \wh\gamma S_x^2 S_y^2
    \wh\gamma S_x S_y  \Big)^{1/2}\no\\
& =&\| \gamma \|_{\cH^{1, (k+1)}}\;,
\eey
where we used the cyclicity (\ref{rootcycle}) several times.

The second term in (\ref{eq:threeterm}) is estimated as
\bey
    \Tr \;\Big| \wt VP_{x, j}\wh\gamma S_x \Big|
   & =&\Tr\;
    \Big( S_x \wh\gamma P_{x, j}\wt V^2 P_{x, j}\wh\gamma S_x\Big)^{1/2}\no\\
  & \leq&\Tr\; \Big( S_x \wh\gamma P_{x, j} S_y^2
    P_{x, j}\wh\gamma S_x\Big)^{1/2}\no\\
   & \leq& \Tr \;\Big( S_x \wh\gamma S_x^2 S_y^2\wh\gamma S_x\Big)^{1/2}
   =\Tr\; \Big( S_xS_y \wh\gamma S_x^2 \wh\gamma S_xS_y\Big)^{1/2}\no\\
   & \leq&\Tr \Big( S_xS_y \wh\gamma S_x^2 S_y^2\wh\gamma S_xS_y\Big)^{1/2}
    \no\\
   & =&\| \gamma \|_{\cH^{1, (k+1)}} \; . \no
\eey
The estimate of the  third term in (\ref{eq:threeterm})
is identical just the second inequality (\ref{eq:Whardy}) is used.
 This completes the proof of Lemma \ref{lemma:hardy}. \myendproof

\section {Removing  the Coulomb cutoffs}\label{sect:cutoff}
\setcounter{equation}{0}

\noindent
{\it Proof of Lemma \ref{lemma:cutoff}}: We fix $N, \delta, \e$ and
for simplicity let $\Psi_t: =  \Psi_{N,t}^\delta$,
$\wt\Psi_t : =  \Psi_{N,t}^{\delta,\e}$. Clearly $\Psi_0=\wt\Psi_0$.
We compute
\bey
    \partial_t \| \Psi_t -\wt\Psi_t\|^2
    &=& 2\mbox{Re} \; \langle  \Psi_{t}
    -\wt\Psi_{t}, W\wt\Psi_{ t}\rangle\no\\
    &\leq & 2\|  \Psi_{t} -\wt\Psi_{t}\| \;
     \|W   \wt\Psi_{ t}\|\; . \no
\eey
We need to estimate
\bey
    \lefteqn{ \|W   \wt\Psi_{ t}\|^2} \label{eq:cuto}\\
    &=& \frac{1}{N^2} \frac{N(N-1)}{2}\int |V_2(x_1-x_2)|^2
    |\wt\Psi(x_1, \ldots x_N)|^2 \rd x_1\ldots
    \rd x_N\no\\
   && + \frac{1}{N^2} \frac{N(N-1)(N-2)}{2} \no\\
    &&\times \int |V_2(x_1-x_2)||V_2(x_2-x_3)|
    |\wt\Psi(x_1, \ldots x_N)|^2 \rd x_1\ldots
    \rd x_N \no\\
   && +  \frac{1}{N^2}\frac{N(N-1)}{2}  \frac{(N-2)(N-3)}{2}\no\\
   && \times \int |V_2(x_1-x_3)||V_2(x_2-x_4)|
    |\wt\Psi(x_1, \ldots x_N)|^2 \rd x_1\ldots
    \rd x_N\no\\
   & \leq&  \int |V_2(x_1-x_2)|^2
    |\wt\Psi(x_1, \ldots x_N)|^2 \rd x_1\ldots
    \rd x_N\no\\
  &&  + \! CN \!\!
    \int \! |V_2(x_2-x_3)|
    \Big( |\nabla_{x_1}\wt\Psi(x_1, \ldots x_N)|^2 \!
    + |\wt\Psi(x_1, \ldots x_N)|^2 \Big)\rd x_1\ldots
    \rd x_N \no\\
   && +  N^2
    \int |V_2(x_1-x_3)||V_2(x_2-x_4)|
    |\wt\Psi(x_1, \ldots x_N)|^2 \rd x_1\ldots
    \rd x_N \; , \no
\eey
where we used that $|V_2(x_1-x_2)| \leq
 |x_1-x_2|^{-1} \leq C(I-\Delta_{x_1})$
for any $x_2$ fixed in the second term.

It follows from  Lemma \ref{lemma:H11}
that
\begin{equation}
     \frac{ \chi(|x-y|\leq \lambda)}{|x-y|^\kappa}
    \leq  C(\kappa)\lambda^{3-\kappa} (I-\Delta_x)(I-\Delta_y)
\label{eq:H11}
\end{equation}
as operators on $L^2(\bR^3\times \bR^3)$ for
$\lambda <1$,  $0\leq \kappa < 3$. We use this estimate  with
$\kappa =1, 2$, and $\lambda = \e/\sqrt{N}$
to continue the estimate as
\bey
    \|W   \wt\Psi_{ t}\|^2 & \leq& C\frac{\e}{\sqrt{N}}
     \Bigg( \| \nabla_{x_1}\nabla_{x_2}\wt\Psi \|^2
    + \|\wt \Psi\|^2\Bigg) \no\\
  &&  + C\e^2\Bigg(
    \| \nabla_{x_1}\nabla_{x_2} \nabla_{x_3}\wt\Psi \|^2
    +  \| \nabla_{x_1} \wt\Psi \|^2 +
    \| \nabla_{x_2} \nabla_{x_3}\wt\Psi \|^2
    +  \|  \wt\Psi \|^2
    \Bigg) \no\\
  &&  +C\e^4\Bigg( \| \nabla_{x_1}\nabla_{x_2}
    \nabla_{x_3}\nabla_{x_4}\wt\Psi \|^2
    + \| \nabla_{x_1}\nabla_{x_2}\wt\Psi \|^2
    + \|\wt \Psi\|^2\Bigg) \no\\
   & \leq& C\e^2 \|\gamma^{\delta, \e, (4)}_{N, t}\|_{\cH^{1, (4)}}\no
\eey
for large enough $N$.
Using (\ref{eq:apr})
this completes the proof of Lemma \ref{lemma:cutoff}. \myendproof

\bigskip

\section{Appendix on partial traces and density matrices}

\setcounter{equation}{0}

In this Appendix we collect a few facts about
trace class operators. We start with the analogue of
the monotone and dominated convergence theorems.

\begin{lemma}\label{lemma:lebesgue}
Let $0\leq A_n \leq A$ be self-adjoint operators
defined on a common separable Hilbert space.

\smallskip

(i) If the sequence is monotone, $A_n\leq A_{n+1}$, $A_n\to A$
strongly, $A$ is compact and $\sup_n \Tr\; A_n <\infty$, then
$\Tr \;  A =\sup_n \Tr\; A_n$, in particular $A$ is trace class.

\smallskip

(ii) If $A_n\to 0$ strongly
and $\Tr \; A <\infty$, then $\Tr \; A_n \to 0$.
\end{lemma}

\noindent
{\it Proof:} Let $\{ f_j\}$ be an eigenbasis for $A$.
For part (i), we have for any $M$ that
$$
    \sum_{i=1}^M  \langle f_i, A f_i\rangle
    = \lim_{n\to\infty} \sum_{i=1}^M  \langle f_i, A_n f_i\rangle
    \leq \sup_n \Tr\; A_n \; .
$$
Taking the limit $M\to\infty$ we see that $\Tr \;  A \leq\sup_n \Tr\; A_n$.
The other direction is trivial.

For part (ii), we notice that $A_n$ is trace class and
$$
    \Tr \; A_n =\sum_{i=1}^\infty  \langle f_i, A_n f_i\rangle\leq
    \sum_{i=1}^N \langle f_i, A_n f_i\rangle + \sum_{i=N+1}^\infty
    \langle f_i, A f_i\rangle \;.
$$
The second term is smaller than any given $\eta>0$ if $N$ is big
enough since $\Tr \; A <\infty$,
 while the first term goes to zero for any fixed $N$ as
$n\to\infty$. \myendproof

\bigskip

Next we show the following  properties of the trace of
the square root.

\begin{lemma}\label{lemma:traces}
Let $A, B$ be positive self-adjoint operators on a common Hilbert
space. Then
\begin{equation}
    \Tr \; \sqrt{ AB^2 A} = \Tr\; \sqrt{BA^2B}
\label{rootcycle}
\end{equation}
and
\begin{equation}
    \Tr \;\sqrt{A+B} \leq 2\Big(\Tr \;\sqrt{A} + \Tr \;\sqrt{B}\Big)\; .
\label{sqr}
\end{equation}
\end{lemma}

\noindent
{\it Proof.} We can assume that at least one side of (\ref{rootcycle}),
say $\Tr\sqrt{ AB^2 A}$ is finite. Then $X= AB^2A\ge 0$ is trace class
and so is $Y:= BA^2B$ by cyclicity of the trace. By functional calculus
$$
    \sqrt {X} = (const.) \int_0^\infty t^{-1/2} X e^{-tX} \rd t \; .
$$
Since $X$ is bounded, the operator $X e^{-tX}$
can be expanded into convergent power series for any $t\ge 0$.
Using the cyclicity of the trace
$$
    \Tr\; X^n = \Tr \;(AB^2A)^n = \Tr\;(BA^2B)^n = \Tr\; Y^n
$$
for any $n$, therefore
$$
    \Tr \; X e^{-tX} = \Tr \; Y e^{-tY}
$$
for any $t\ge 0$ and (\ref{rootcycle}) follows after integration.
We used Lemma \ref{lemma:lebesgue} (i) to interchange the trace
and the improper $\rd t$ integration.

\medskip

For the proof of (\ref{sqr}) we let $\lambda_1(A)\ge \lambda_2(A)\ge
\ldots \ge 0$ be the eigenvalues of $A$, similar notation
is used for $B$ and $A+B$.
Fan's inequality \cite{Si} asserts that
$$
    \lambda_{n+m+1} (A+B) \leq \lambda_{n+1}(A) + \lambda_{m+1}(B)
$$
for any $m,n\ge 0$. Therefore
\bey
    \lambda_{2n+1}(A+B)&\leq& \lambda_{n+1}(A) + \lambda_{n+1}(B) \no\\
    \lambda_{2n}(A+B)&\leq& \lambda_{2n-1}(A+B)
    \leq \lambda_n(A) + \lambda_n(B)\no
\eey
and
\bey
    \Tr\;\sqrt{A+B} &=&\sum_{k=1}^\infty \sqrt{\lambda_k(A+B)}\no\\
    &\leq& 2\sum_{n=1}^\infty \sqrt{\lambda_{n}(A) + \lambda_{n}(B)}\no\\
    &\leq& 2\sum_{n=1}^\infty
    \Big(\sqrt{\lambda_{n}(A)} +  \sqrt{\lambda_{n}(B)} \Big)\no\\
    &=& 2 \Big( \Tr\;\sqrt{A}  + \Tr\; \sqrt{B}\Big)
     \; .\qquad\qquad\no
\eey
\myendproof

\bigskip

Now we give the  definition
of the partial trace. For more details, see e.g., \cite{Par}.

\begin{definition} \label{def:partial}
Let $H_1, H_2$ be separable Hilbert spaces.
Let $A$ be a trace class operator on the tensor product
$H=H_1\otimes H_2$. Then there exists a unique
operator $B$ in $\cL^1(H_1)$ such that
\begin{equation}
    \Tr_{H_1}\big[ BK\big] = \Tr_{H}\big[ A(K\otimes I)\big]
\label{def:partialtr}
\end{equation}
for every $K\in \cK(H_1)$. This operator $B$ is called
the partial trace of $A$ with respect to $H_2$
and is denoted by $\Tr_{H_2} A$ or just by $\Tr_2 \; A$.
\end{definition}
The existence of the partial trace follows from
duality: the linear functional $K\mapsto \Tr_{H} A(K\otimes I)$
defines a bounded linear map on $\cK(H_1)$
and $[\cK(H_1)]^*=\cL^1(H_1)$.

\begin{proposition}\label{prop:parttr}
The partial trace satisfies the following relations
\begin{equation}
    \Tr_{H_1} \big| \Tr_{H_2} (I\otimes A)B\big| =
    \Tr_{H_1} \big| \Tr_{H_2} B (I\otimes A) \big|
\label{partcycle}
\end{equation}
and
\begin{equation}
    \Tr_{H_1} \big| \Tr_{H_2} A\big| \leq
    \Tr_{H} |A|\;.
\label{fubini}
\end{equation}
\end{proposition}

\noindent
{\it Proof}: For the proof of (\ref{partcycle})
 we use the variational principle (\ref{eq:var})
\bey
    \Tr_{1} \big| \Tr_{2} (I\otimes A)B\big|
    &=& \sup  \Tr_1 \; K \big(
     \Tr_{2} (I\otimes A)B\big) \no\\
    &=& \sup  \Tr_{1,2} \;
     (K\otimes I) (I\otimes A)B \no\\
    &=& \sup \Tr_{1,2} \;
    (K\otimes I)B (I\otimes A)   \no\\
    &=&  \sup  \Tr_1 \; K \big(
     \Tr_{2} B(I\otimes A)\big) \no\\
    & = & \Tr_{1} \big| \Tr_{2} B (I\otimes A) \big|
\eey
where the supremum is over all $K\in \cK(H_1)$ with $\|K \|=1$.
We used the cyclicity of $\Tr_{1,2}$ and that $I\otimes A$ commutes
with $K\otimes I$.

\bigskip

For the proof of (\ref{fubini}) we first
 observe that the variational principle
(\ref{eq:var}) extends to bounded operators as follows:
\begin{equation}
    \| A\|_1
%    \; \sup_{K\in \cK(H)\; : \; \| K \|=1} \Big| \; \Tr \; AK\;
%     \Big| =
   = \sup_{L\in \cL(H)\; : \; \| L \|=1} \Big| \; \Tr \; AL\;
     \Big| \;
\label{eq:varb}
\end{equation}
for any $A\in \cL^1(H)$. The proof follows from $\cK(H)\subset \cL(H)$
on one hand, and from $|\Tr AL |\leq \| A \|_1 \|L\|$
on the other hand, using (\ref{eq:AB}).

\medskip

\noindent
Therefore
\bey
    \Tr_{1}\;  \big| \Tr_{2}\; A\big| &=&
      \sup_{L\in \cL(H_1) \; : \;\|L \|=1} \Tr_{1} \; L \; \big[\Tr_{2}\; A
    \big] \no\\
    &=& \sup_{L\in \cL(H_1) \; : \; \|L \|=1}
    \Tr_{1,2} \; (L\otimes I) A \no\\
    &\leq& \Tr_{1,2}\; |A| \; . \no
\eey
The estimate follows again from the variational principle (\ref{eq:varb})
since $L\otimes I \in \cL(H_1\times H_2)$ and $\| L\otimes I\|= \|L\|$.
 \myendproof

\bigskip

Finally, we prove that the space $\cA$ is indeed the  dual space
of the Sobolev space $\cH^1$:

\medskip

\noindent
{\it Proof of Lemma \ref{lemma:dual}}.
(i) $\cH^1\subset \cA^*$. Let $T\in \cH^1$ and
we write any element $A\in \cA$ as $A=SKS$ with some $S\in \cK$.
Then
$$
    \Big| \Tr\; TA \Big| = \Big| \Tr \; TSKS \Big| \leq \| K \|\;
    \Tr\;  |STS| = \| SKS \|_\cA \; \|T\|_{\cH^1} =
     \| A \|_\cA \; \|T\|_{\cH^1}
$$
which shows that $\ell_T: A \to \Tr\;  TA $ is a continuous linear functional
on $\cA$ indexed by $T\in \cH^1$ and $\| \ell_T \|_{\cA^*}\leq \|T\|_{\cH^1}$.

\medskip

(ii) $\cA^*\subset\cH^1$. Let $f\in \cA^*$ be a continuous linear
functional on $\cA$. Consider $\psi, \phi \in H^1(\bR^3)\subset H$,
i.e., $S\psi, S\phi \in H$.  Since $|\phi\rangle\langle\psi|$ is
compact,  $|S\phi\rangle\langle S\psi|
= S |\phi\rangle\langle\psi| S$ is contained in $\cA$.
Consider the sesquilinear functional
$$
    \cF_f:
    (\psi, \phi)\mapsto f\Big[ \; S |\phi\rangle\langle\psi| S\; \Big]
$$
from $H^1\times H^1 \to \bC$. This sesquilinear map is continuous since
$$
    \Bigg| f\Big[ \; S |\phi\rangle\langle \psi| S\; \Big] \Bigg|
    \leq \| f\|_{\cA^*} \Big\|  \; S |\phi\rangle\langle\psi| S \;
     \Big\|_{\cA}
    = \| f\|_{\cA^*} \Big\| \;  |\phi\rangle\langle\psi| \; \Big\|
    \leq \| f\|_{\cA^*} \| \phi\| \; \|\psi\| \;,
$$
moreover it  clearly extends from  $H^1\times H^1$ to $H\times H$.

Hence there exists a unique bounded operator $B= B_f$ such that
$$
    (\psi, B\phi) =  f\Big[ \; S |\phi\rangle\langle\psi| S\; \Big]
$$
for every $\phi,\psi\in H$, and $\| B \|\leq \| f\|_{\cA^*}$
(uniqueness follows from the density of $H^1\subset H$).

By polar decomposition we write $B=U |B|$ with some partial isometry $U$.
Let $\{ \psi_i\}$ be a finite orthonormal set, we can write
\bey
    \sum_{i=1}^N (\psi_i, |B| \psi_i)
    &= &f \Big[ \; S
    \Big(\sum_i |\psi_i\rangle\langle U\psi_i|\Big) S\; \Big] \no \\
   & \leq&  \| f\|_{\cA^*} \Big\| \sum_i |\psi_i\rangle\langle U\psi_i|
    \Big\| \no\\
    &=& \| f\|_{\cA^*} \;.
\label{eq:trace}\eey
Here we used that $U\psi_i$ is also orthonormal, and for any $h\in H$
\bey
    \Big( h,  \sum_i |\psi_i\rangle\langle U\psi_i| h\Big)
    &=& \sum_i (h,\psi_i)(U\psi_i, h) \no\\
&\leq &
    \frac{1}{2}\sum_i  \Big[ |(h,\psi_i)|^2 + |(U\psi_i, h)|^2\Big] \no\\
    &\leq& \| h\|^2 \;. \no
\eey
Hence taking the supremum for all orthonormal sets in (\ref{eq:trace}),
we see that $\Tr \; |B|\leq \| f\|_{\cA^*}$.

Finally, we define $T_f : = S^{-1}B_fS^{-1}$ for any $f\in \cA^*$.
Then clearly
$$
    \| T_f \|_{\cH^1} = \Tr \; |B_f|\leq \| f\|_{\cA^*}
$$
and for any $A\in \cA$
$$
    \Tr \; (T_fA) = f(A)
$$
since this is valid for all $A= S \sum_{i=1}^n
 |\phi_i\rangle\langle \psi_i| S$
finite range operators and these are dense
in $(\cA, \| \cdot \|_{\cA})$.
Hence for any $f\in \cA^*$ we found a representative $T_f\in \cH^1$
with smaller or equal norm. \myendproof

\end{document}